\documentclass[%
 reprint,
 amsmath,amssymb,
 aps,
]{revtex4-2}

\usepackage{graphicx}% Include figure files
\usepackage{dcolumn}% Align table columns on decimal point
\usepackage{bm}% bold math
\begin{document}
%TC:ignore
\title{Three Dimensional Reconfigurable Optical Singularities in Bilayer Photonic Crystals}

\author{Xueqi Ni$^1$}
\author{Yuan Liu$^1$}
\author{Beicheng Lou$^2$}
\author{Mingjie Zhang$^1$}
\author{Evelyn L. Hu$^1$}
\author{Shanhui Fan$^2$}
    \email{shanhui@stanford.edu}
\author{Eric Mazur$^1$}
    \email{mazur@seas.harvard.edu}
\author{Haoning Tang$^1$}
    \email{hat431@g.harvard.edu}

\address{$^1$ School of Engineering and Applied Sciences, Harvard University, Cambridge, MA 02138, USA}
\address{$^2$ Department of Applied Physics and Ginzton Laboratory, Stanford University, Stanford, CA 94305, USA}

\begin{abstract}
Metasurfaces and photonic crystals have revolutionized classical and quantum manipulation of light, and opened the door to studying various optical singularities related to phases and polarization states. However, traditional nanophotonic devices lack reconfigurability, hindering the dynamic switching and optimization of optical singularities. This paper delves into the underexplored concept of tunable bilayer photonic crystals (BPhCs), which offer rich interlayer coupling effects. Utilizing silicon nitride-based BPhCs, we demonstrate tunable bidirectional and unidirectional polarization singularities, along with spatiotemporal phase singularities. Leveraging these tunable singularities, we achieve dynamic modulation of bound-state-in-continuum states, unidirectional guided resonances, and both longitudinal and transverse orbital angular momentum. Our work paves the way for multidimensional control over polarization and phase, inspiring new directions in ultrafast optics, optoelectronics, and quantum optics.
\end{abstract}
%TC:endignore

\maketitle

%\paragraph{Introduction}
Multidimensional optical singularities offer promising solutions for the exponentially growing information capacity in optical communication and quantum manipulation\cite{RN74,RN11,RN70,RN14,RN12}. These singularities manifest themselves as undefined phases or polarizations in momentum or frequency spaces, with unique topological properties and optical characteristics\cite{RN47,RN61,RN73,RN55,RN68,RN63,RN64,RN52,RN58}. Research in this area is garnering considerable attention in recent years due to the potential applications of these optical singularities. For instance, vortex-like polarization singularities (V points) eliminate far-field radiation, leading to bound states in the continuum (BIC)\cite{RN30,RN72,RN51,RN54,RN53,RN49} and unidirectional guided resonances (UGR)\cite{RN27,RN4,RN56}. Spiral-shaped phase singularities in momentum and momentum-frequency space enable beams to carry unbounded orbital angular momentum (OAM)\cite{RN50, RN69,RN48,RN8,RN66,RN65}. The manipulation of the directionality, polarization state, and angular momentum of photons, introduces novel degrees of freedom for optical control. This enables applications across classical and quantum regimes, encompassing areas such as lasing\cite{RN38,RN48, RN40,RN77}, spatiotemporal modulation\cite{RN67}, ultra-low-loss communication\cite{RN26}, single-photon quantum sources\cite{RN10}, and hybrid quantum states\cite{RN11,RN12}.

Although traditional benchtop optical systems can generate various optical singularities, photonic crystals, metasurfaces, and micro spiral phase plates offer advantages such as lower power consumption and ease of integration for practical applications\cite{Rn75,RN67,RN8, RN7, RN46, RN45, RN39, RN57, RN44}. Metasurfaces and spiral phase plates use complex designs in real space to produce singularities; photonic crystals exploit resonant modes in momentum space topologies to generate polarization and phase singularities. In contrast to metasurfaces and spiral phase plates, photonic crystals do not require a geometric center and are easily manufactured. However, like all single-layer devices, they cannot be reconfigured once fabricated. Bilayer photonic crystals (BPhCs)\cite{RN24, RN6,RN20,RN16,RN25,RN28,PhysRevLett.131.053602,RN22,RN62,RN33,RN36,RN23,RN17,RN9,PhysRevB.107.155402,saadi2023supercells}, on the other hand, possess additional degrees of freedom, such as twist angle, in-plane displacement, and interlayer spacing, which allow tuning of optical singularities and manipulating optical properties. Moreoever, MEMS-nanophotonic technology enables dynamic adjustment of these singularities and properties\cite{RN31, RN29,RN43}.

In this paper, we study the mechanisms for controlling momentum and momentum-frequency singularities in tunable BPhCs. We will focus on translated rather than twisted bilayer structures because translation is easier to accomplish on chip than rotation. In particular, we will show that such translationally-tunable BPhCs can be used to adjust: 1. bidirectional polarization singularities and BICs; 2. unidirectional polarization singularities and UGRs; 3. spatiotemporal phase singularities and transverse OAM. Our results offer new insights into manipulating multidimensional optical singularities in microscale devices. 

\begin{figure}[h]
\includegraphics[width = .5\textwidth]{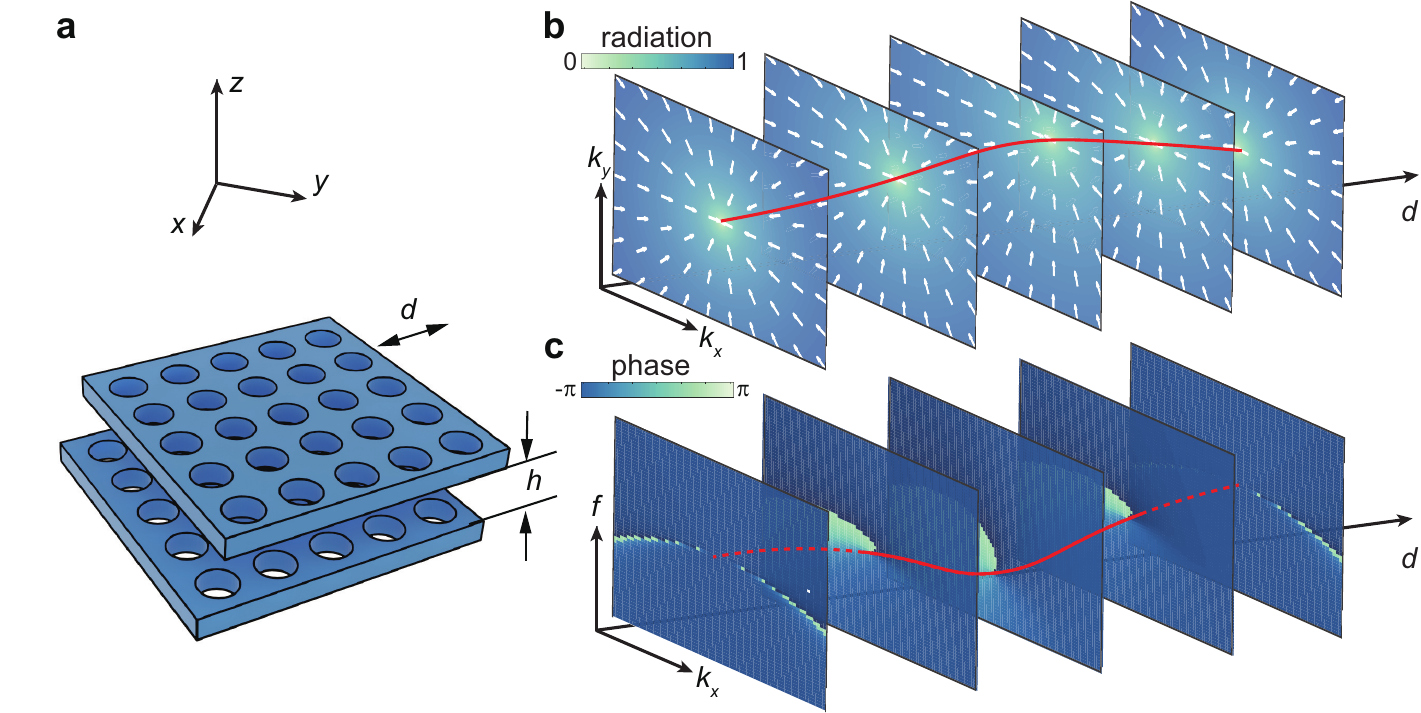}
\caption{Concept of three-dimensional optical singularities in bilayer photonic crystals.  (a) Schematic of bilayer photonic crystal. $h$ and vector $\mathbf{d}$ represent gap size and relative displacement between two photonic crystal slabs. (b) Polarization ellipses in the far field and corresponding radiation strengths in momentum space. (c) Frequency-momentum space phase singularities. Slices are taken at different $\mathbf{d}$.}
\end{figure}

%\paragraph{Results}
The individual photonic crystal slabs that comprise the BPhCs are made of SiN ($n$ = 2.02) and have $C_{4v}$ symmetry and circular airholes. The radius of the airholes is $r = 0.33a$, where $a$ is the length of a unit cell, and the thickness of each slab is $0.54a$. Bilayer photonic crystal slabs are separated by an interlayer gap $h$, and misaligned by an in-plane vector $\mathbf{d}$ (Fig. 1a). These two geometric parameters, $\mathbf{d}$ and $h$, permit modulating the positions of the polarization and phase singularities along continuous nodal lines in momentum space (Fig. 1b) and frequency-momentum space (Fig. 1c). In the following sections, we will elaborate on the mechanism for modulating optical singularities using bilayer photonic crystals.

\begin{figure}[htbp]
    \includegraphics[width= .5\textwidth]{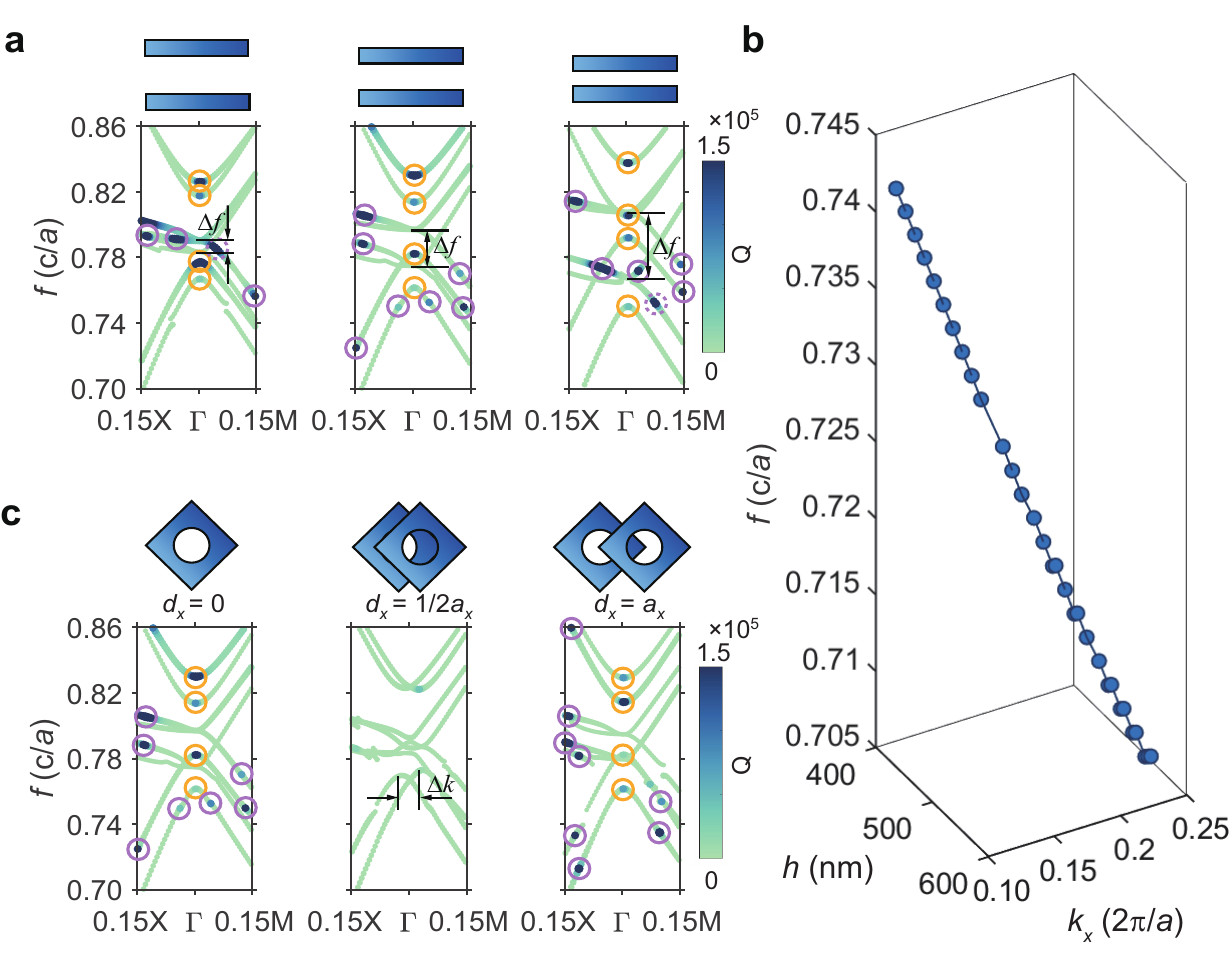}
    \caption{Evolution of band structures for bilayer photonic crystal slabs.  (a) Band structures for bilayer photonic crystal slabs when the interlayer gap $h$ varies. $\Delta f$ represents the frequency variation between symmetric mode and anti-symmetric mode. Yellow circles indicate the symmetry-protected BICs. Purple circles indicate the accidental BICs. (b) Frequency and gap $h$ dependency of an accidental BIC in momentum space. (c) Band structures for bilayer photonic crystal slabs when the in-plane shift $\mathbf{d_x}$ varies at fixed $h$ = 600nm and $d_y$=0. $\Delta k$ represents the separation of the band edge. The color bar represents quality factors of modes. }
    \end{figure}
We begin by exploring the band structure of BPhCs and describe a class of polarization singularities, known as bound states in the continuum. BICs exhibit infinite $Q$-factors, signifying complete energy localization within the structure. In monolayer $C_{4v}$ photonic crystals, near the $\Gamma$ point, the lowest-frequency guided resonances consist of four modes, corresponding to the four lowest-order in-plane reciprocal wavevectors\cite{RN21, PhysRevB.65.235112}. The bilayer band structure comprises two sets of these fundamental modes: symmetric and anti-symmetric modes, exhibiting either in-phase or out-of-phase electric fields within the two layers (Supplementary Information)\cite{RN60, RN5}. As shown in Fig. 2a, the frequency difference between the two sets of modes, denoted as $\Delta f$, is determined by the coupling strength between the two PhC layers. In BPhCs, two types of BICs exist: accidental BICs (purple circles), which can be achieved by tuning structural parameters, and symmetry-protected BICs (yellow circles), limited to specific high-symmetry points due to a symmetry mismatch between photonic crystal modes and far-field plane waves. Altering $h$ modulates $\Delta f$, as well as the positions of BICs. When the distance between two layers of the photonic crystals is relatively large ($h > \lambda$), the interlayer coupling is weak and the two sets of modes approach degeneracy. Notably, with changing $h$, symmetry-protected BICs move only in frequency space and remain restricted to the $\Gamma$ point, while accidental BICs not only separate in frequency space but also change in momentum space. Fig. 2b illustrates the movement of a set of accidental BIC-associated polarization singularities in frequency and momentum space with varying $h$.

Next, we explore the effects of keeping $h$ fixed and varying $\mathbf{d}$, which changes the in-plane coupling of the photonic crystal slabs. We decompose the displacement $\mathbf{d}$ into x and y components ($\mathbf{d} = d_x\hat{\mathbf{x}} + d_y\hat{\mathbf{y}}$) and normalize it to the unit cell length $|\mathbf{a}|$, where $\mathbf{a} = a_x\hat{\mathbf{x}}+a_y\hat{\mathbf{y}}$ is the lattice vector pointing from (0,0) to $(a/\sqrt{2}, a/\sqrt{2})$. The variation of $\mathbf{d}$ gives rise to an additional phase term $exp(ik_xd_x+ik_yd_y)$ in the in-plane coupling term. In terms of the band structure, the change in intralayer coupling manifests itself as a displacement $\Delta k$ of the bandedge in momentum space, while barely affecting the mode frequency splitting $\Delta f$ (Fig.2c). Furthermore, $\mathbf{d}$ breaks the up-down mirror symmetry and inversion symmetry (except 
 under high symmetric configurations such as $\mathbf{d} = a/\sqrt{2} \hat{\mathbf{x}}$), leading to the disappearance of the symmetry-protected BICs at $\Gamma$\cite{10.1063/1.1563739} and the accidental BICs at off-$\Gamma$ points. However, due to the topological charge conservation of far-field polarizations, the polarization singularities corresponding to BICs do not vanish. Instead, they relocate in momentum space or merge with other singularities. A consequence of breaking the up-down mirror symmetry is that the upward and downward radiation fields from the BPhC become different. As a result, the bidirectional polarization singularities associated with BICs evolve into unidirectional ones. In such scenarios, the high $Q$-factor no longer serves as an indicator of polarization singularity. A detailed analysis of polarization fields in momentum space, considering both upward and downward radiations, becomes necessary to understand the radiation pattern.

\begin{figure*}[htbp]
    \includegraphics[width = \textwidth]{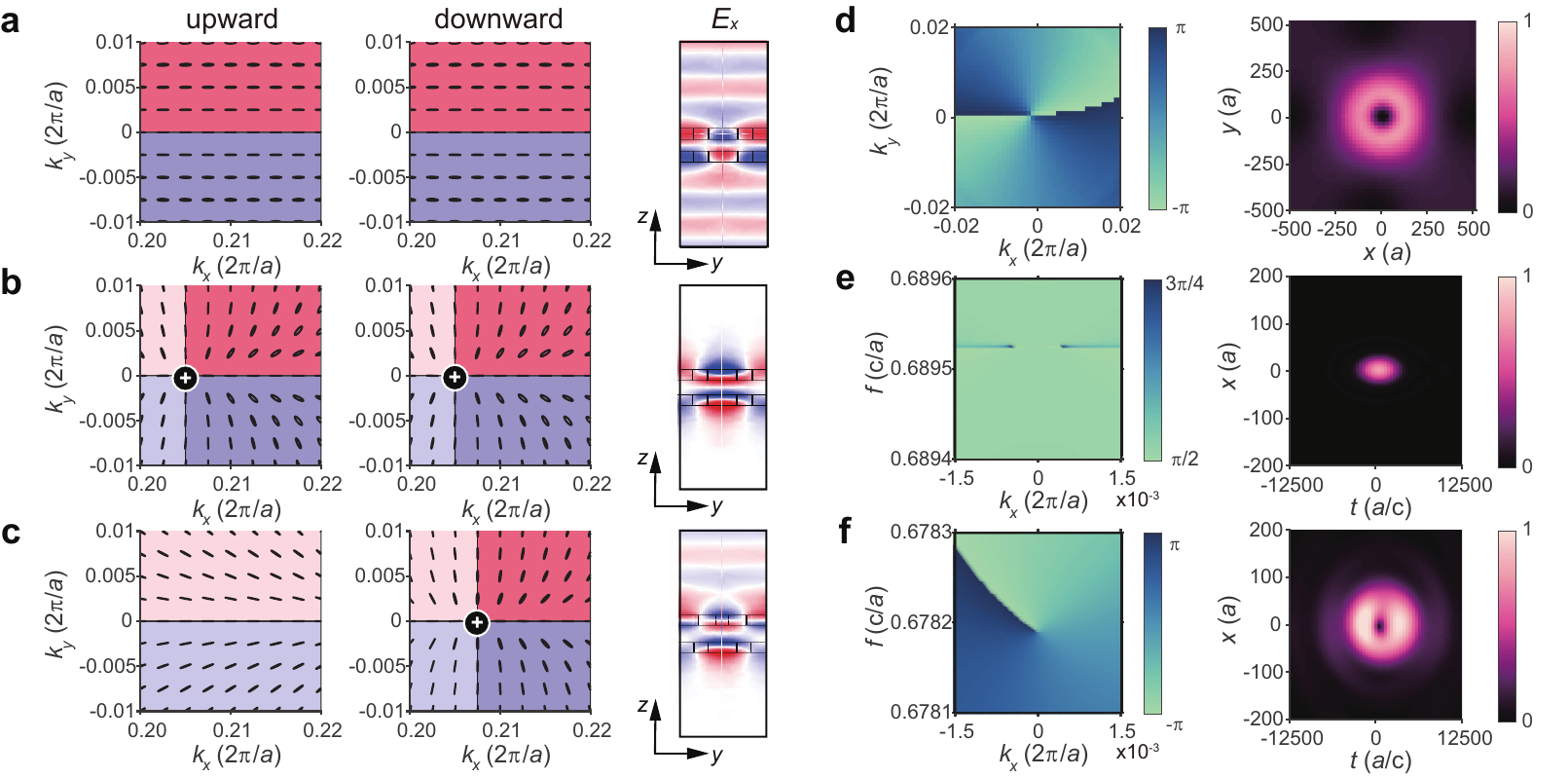}
    \caption{Three-dimensional control of optical radiation.  (a-c) Upward (left) and downward (middle) far-field polarization states in momentum space, and the corresponding cross-sectional electric field  (right) for (a) radiative modes, (b) BICs, and (c) UGRs. Geometric parameters for (a-c) are $h = (450\mathrm{~nm}, 550\mathrm{~nm}, 630\mathrm{~nm})$, $d_x = (0, 0, 0.3a_x)$, $d_y = (0, 0, 0)$. Different regions are shaded according to the signs of $E_x$ and $E_y$. (d-f) Phase (left) and intensity (right) of electromagnetic field excited by a Gaussian pulse for (d) spatial vortex beam with longitudinal OAM, (e) in the absence of transverse OAM, and (f) spatiotemporal vortex beam with transverse OAM. Geometric parameters for (d-f) are $h = (600\mathrm{~nm}, 600\mathrm{~nm}, 600\mathrm{~nm})$, $d_x = (0, 0, 0.45a_x)$, $d_y = (0, 0, 0)$.}
\end{figure*}
In Fig.3a-c, we demonstrate the unidirectionality of the singularities according to the polarization fields along the out-of-plane $+z$ and $-z$ directions. These singularities, arising from topological defects in momentum space\cite{RN13}, can be described by their topological charges:
\begin{equation}
    q = \frac{1}{2\pi}\oint_C d\mathbf{k}\cdot \nabla_k \phi(\mathbf{k})
\end{equation}
where $\phi(\mathbf{k})$ is the azimuthal angle of polarization major axis, and C is a closed loop in momentum space. When there are no polarization singularities, all of momentum space has well-defined polarization states, resulting in a trivial topological charge ($q = 0$). Consequently, the electric field of the BPhC radiates outward on both sides (Fig.3a). When the displacement between BPhCs is $\mathbf{d} = \mathbf{0}$, implying up-down mirror symmetry, bidirectional polarization singularities emerge at $\Gamma$ points or at off-$\Gamma$ points by adjusting $h$, as illustrated in Fig.2. Around these singularities, far-field polarization states on both sides of the BPhC possesses nontrivial topological charges ($q = 1$), leading to high-$Q$ resonances due to the absence of radiative losses (Fig. 3b). Furthermore, when the up-down mirror symmetry is broken, the  upward and downward polarization singularities shift in opposite directions in momentum space. As a result, the integration over a closed loop gives rise to different topological charges on each side of the BPhC. The appearance of these single-sided polarization singularities leads to unidirectional guided resonances (UGRs), where the guided resonance radiates only towards one side of the BPhC (Fig. 3c). The directionality of the radiation and the position of the unidirectional polarization singularity in momentum space is determined by the displacement $\mathbf{d}$ (Supplementary Information).

In addition to manipulating the directionality of polarization singularities, BPhCs can also generate orbital angular momentum. Through spin-orbit coupling, polarization singularities can be transformed into phase singularities\cite{RN59, RN18,RN41}. For instance, by directing a circularly polarized Gaussian pulses towards the BPhC, cross-polarized output light with wave vectors near the singularity carry a helical geometric phase, also known as Pancharatnam-Berry phase. The helical phase results in a spatial optical vortex and longitudinal orbital angular momentum (Fig.3d). Furthermore, a non-local transformation of longitudinal and transverse orbital angular momentum can be achieved by changing the photonic crystal symmetry. We numerically simulated the intensity distribution of the optical field in the space-time ($x$-$t$) domain after the incident pulse passes through the BPhC. When the displacement between two layers is zero, $d_x = 0$, phase singularities only occur in momentum space, with a characteristic parabolic phase distribution in the frequency-momentum ($f$-$k$) space. The absence of isolated singularities in the $f$-$k$ space prevents the formation of spatiotemporal optical vortices (Fig.3e). 

When $d_x \neq 0$, a topologically protected isolated singularity emerges in the $f$-$k$ space as a result of breaking the up-down mirror symmetry. The $f$-$k$ singularity also represents a helical phase distribution that modulates the transmitted pulse, as depicted in the left column of Fig.3f. In contrast to the momentum space singularities, the $f$-$k$ singularity and the corresponding helical phase yield a distinctive donut-shaped pattern in spatiotemporal space, with zero amplitude at the center of the donut. The spatiotemporal donut-shaped field means the BPhC carries transverse orbital angular momentum\cite{RN67}. The manipulation of longitudinal and transverse orbital angular momentum has applications in optical communication, ultrafast optics, and light-matter interactions that require multiple degrees of freedom to adjust.

\begin{figure}[htbp]
    \includegraphics[width = .5\textwidth]{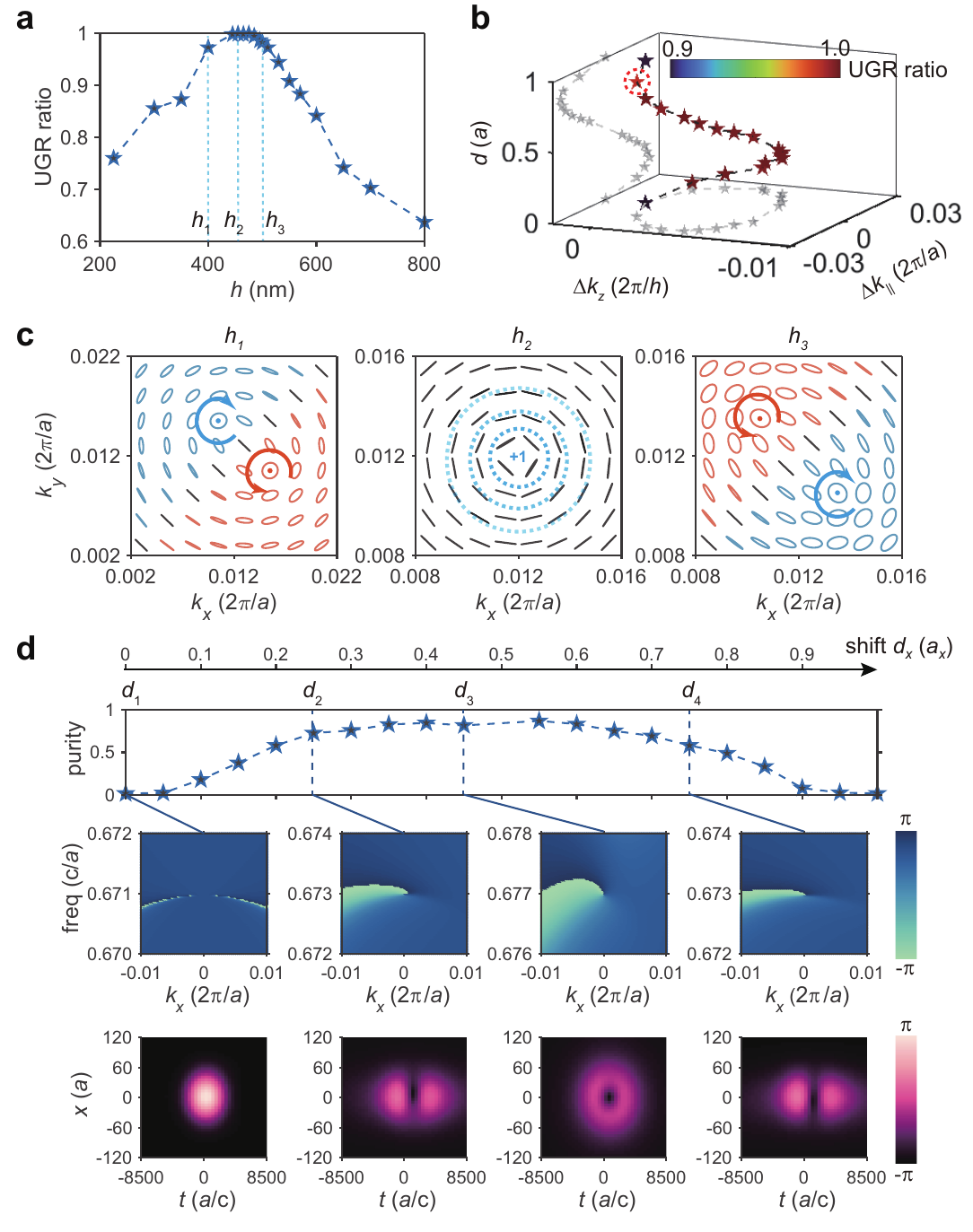}
    \caption{Reconfigurability of polarization and phase singularties. (a) UGR ratios for fixed displacement $(d_x, d_y) = (0.95a_x, 0.95a_y)$ and varying $h$. (b) 3D trace of UGRs when $h$ and $\mathbf{d}$ vary simultaneously. Colors represent UGR ratios. The red dashed circle labels the displacement corresponding to (a). (c) Polarization fields taken at $h_1$, $h_2$, and $h_3$. (d)Spatiotemporal modulation of light in BPhCs when varying $d_x$, showing calculated OAM purities (upper row), phase diagram in frequency-momentum space for $(d_{1x}, d_{2x}, d_{3x}, d_{4x}) = (0, 0.25, 0.45, 0.75)a_x$ (middle row), and the corresponding transmitted electromagnetic field intensity using a Gaussian input pulse (bottom row). The center of transmitted donut pattern slightly deviates from $t = 0$ due to imperfections of phase distributions.}
\end{figure}
    
 The degrees of freedom, $h$ and $\mathbf{d}$, in translational tunable BPhCs enable versatile adjustments and reconfigurations of unidirectional polarization singularities in momentum space. We can use the ratio $|E_{up} - E_{down}|/|E_{up} + E_{down}|$, where $E_{up(down)}$ is the energy flux in upward (downward) direction, as the UGR ratio, to quantify the unidirectionality of the far-field radiation. When this UGR ratio is 1, we obtain perfect single-sided emission. Fig. 4a reveals optimized UGR ratios for varied $h$ and $k$ at fixed $\mathbf{d}$. Notably, when $\mathbf{d}$ remains constant, only a specific $h = h_2$ yields a UGR ratio of 1. The relatively low UGR ratio for other values of $h$ is due to $C_2$ symmetry breaking, which causes the original vortex polarization singularity with charge $q = 1$ to split into two circularly polarized points (C points) with opposite handedness. Left- and right-handed circularly polarized light couple to different wave vectors when $h = h_{1,3}$, resulting in chiral dichroism in the optical response. Only when $h = h_2$, do the C points merge and form a unidirectional polarization vortex center, as depicted in the middle column in Fig. 4c. The merging behavior shows that the appearance of unidirectional polarization singularities is accidental. The origin of the accidental singularities can be phenomenologically explained by phase cancellation between the in-plane and out-of-plane electromagnetic fields. Specifically, the upward and downward radiation intensities can be represented using a differential equation:
\begin{equation}
\begin{aligned}
\frac{d}{d t}\left(\begin{array}{l}A_1 \\ A_2\end{array}\right)=i\left(\begin{array}{cc}\omega & J \\ J^{\dagger} & \omega\end{array}\right)\left(\begin{array}{l}A_1 \\ A_2\end{array}\right)+\mathbf{D}^{\dagger} \mathbf{D}\left(\begin{array}{l}A_1 \\ A_2\end{array}\right)
\end{aligned}
\end{equation}
where $\omega$ represents the resonance frequency of the mode in each layer, $J$ is the interlayer coupling between the modes in the two layers, $A_{1(2)}$ is the mode amplitude of the first (second) layer, and $\mathbf{D}$ is the coupling matrix between the modes and far-field radiations, taking the form:
\begin{equation}
\begin{aligned}
\mathbf{D}^{\dagger}=\left(\begin{array}{llll}C_{1d s}(\mathbf{k}) & C_{1d p}(\mathbf{k}) & C_{1 u s}(\mathbf{k}) & C_{1 u p}(\mathbf{k}) \\ C_{2 d s}(\mathbf{k}) & C_{2 d p}(\mathbf{k}) & C_{2 u s}(\mathbf{k}) & C_{2 u p}(\mathbf{k})\end{array}\right)
\end{aligned}
\end{equation}
$C_{1(2)u(d)s(p)}$ represents the radiative strength related to the s(p)-polarized component in the upward (downward) space induced by the first (second) layer. Simultaneously tuning $d_x$, $d_y$, and $h$ leads to a change in coupling strength $C_{2 d s}(k)=C_{1 d s}(k) e^{i\left(d_x k_x+d_y k_y+\Delta h k_z\right)}$. Eliminating far-field radiation, $i.e.$, $\mathbf{DA} = \mathbf{0}$, requires: $d_x k_x+d_y k_y+\Delta h k_z = (2n-1)\pi, n\in Z$.

In Fig. 4b, we depict the trajectory of unidirectional polarization singularities between $(d_x, d_y) = (0, 0)$ and $(a_x, a_y)$. For $\mathbf{d} = \mathbf{0}$ and $\mathbf{d} = \mathbf{a}$, we obtain the same up-down mirror symmetric configuration. Therefore, in momentum space, these singularities trace a closed loop. In this example, we have $d_x = d_y$, which implies that the singularities move along the $k_x = k_y$ direction in addition to a shift of $k_z$, and we only show downward UGRs. Other directions are discussed in the Supplementary Information. The displayed UGR ratios exceed 0.99.

In addition to the tunability of unidirectional polarization singularities, tunable BPhCs facilitate dynamic modulation of $f$-$k$ singularities and their corresponding spatiotemporal optical vortices. To demonstrate this modulation, we simulated the field distribution of a Gaussian pulse (centered near the lowest TE bandedge with a waist width of 32$a$) after passing through the BPhC (Fig.4d). We calculated the fields and OAM purities from $(d_x, d_y) = (0, 0)$ to $(a_x, 0)$. The method for calculating OAM purity is discussed in the Supplementary Information\cite{RN37}. As the helical phase becomes more uniformly distributed around the singularity within the range of $-\pi$ to $\pi$, the purity approaches 1, resulting in the generation of higher-quality spatiotemporal optical vortices. In the middle and bottom rows in Fig. 4d, we show the phase and intensity distribution of the transmitted field for four characteristic displacements labeled $d_{1-4}$. At displacement $d_3$, we obtain a purity of 0.8, resulting in a donut pattern in spatio-temporal space. At $d = d_{2,4}$, the phase distributions deviate from the standard helical phase, causing the optical vortex to split along the time axis. Upon the disappearance of the isolated singularity ($d = d_1$), the total topological charge approaches zero, leading to the absence of orbital angular momentum in the outgoing pulse. The ability of tunable BPhCs to adjust the optical singularities in freqency-momentum space opens the door to optimizing spatiotemporal optical vortices at the microscale. 

In conclusion, tunable bilayer photonic crystals offer a high degree of reconfigurability for tailoring the direction, wavelength, polarization, orbital angular momentum of light. We show that by adjusting the interlayer gap $h$, BPhCs allow adjusting bidirectional polarization singularities in momentum space. Further breaking the symmetry by introducing a translational degree of freedom $\mathbf{d}$ opens the door to unidirectional polarization singularities and unidirectional guided resonances.  Bilayer photonic crystals  also permit modulating phase singularities within the frequency-momentum domain and optimizing spatiotemporal optical vortices. Our results are applicable to any waves where tuning multiple degrees of freedom is of interest, including electromagnetic waves across the entire spectrum and mechanical waves.

\bibliography{apssamp}% Produces the bibliography via BibTeX.

\providecommand{\noopsort}[1]{}\providecommand{\singleletter}[1]{#1}%
\begin{thebibliography}{71}%
\makeatletter
\providecommand \@ifxundefined [1]{%
 \@ifx{#1\undefined}
}%
\providecommand \@ifnum [1]{%
 \ifnum #1\expandafter \@firstoftwo
 \else \expandafter \@secondoftwo
 \fi
}%
\providecommand \@ifx [1]{%
 \ifx #1\expandafter \@firstoftwo
 \else \expandafter \@secondoftwo
 \fi
}%
\providecommand \natexlab [1]{#1}%
\providecommand \enquote  [1]{``#1''}%
\providecommand \bibnamefont  [1]{#1}%
\providecommand \bibfnamefont [1]{#1}%
\providecommand \citenamefont [1]{#1}%
\providecommand \href@noop [0]{\@secondoftwo}%
\providecommand \href [0]{\begingroup \@sanitize@url \@href}%
\providecommand \@href[1]{\@@startlink{#1}\@@href}%
\providecommand \@@href[1]{\endgroup#1\@@endlink}%
\providecommand \@sanitize@url [0]{\catcode `\\12\catcode `\$12\catcode `\&12\catcode `\#12\catcode `\^12\catcode `\_12\catcode `\%12\relax}%
\providecommand \@@startlink[1]{}%
\providecommand \@@endlink[0]{}%
\providecommand \url  [0]{\begingroup\@sanitize@url \@url }%
\providecommand \@url [1]{\endgroup\@href {#1}{\urlprefix }}%
\providecommand \urlprefix  [0]{URL }%
\providecommand \Eprint [0]{\href }%
\providecommand \doibase [0]{https://doi.org/}%
\providecommand \selectlanguage [0]{\@gobble}%
\providecommand \bibinfo  [0]{\@secondoftwo}%
\providecommand \bibfield  [0]{\@secondoftwo}%
\providecommand \translation [1]{[#1]}%
\providecommand \BibitemOpen [0]{}%
\providecommand \bibitemStop [0]{}%
\providecommand \bibitemNoStop [0]{.\EOS\space}%
\providecommand \EOS [0]{\spacefactor3000\relax}%
\providecommand \BibitemShut  [1]{\csname bibitem#1\endcsname}%
\let\auto@bib@innerbib\@empty
%</preamble>
\bibitem [{\citenamefont {Ni}\ \emph {et~al.}(2021)\citenamefont {Ni}, \citenamefont {Huang}, \citenamefont {Zhou}, \citenamefont {Gu}, \citenamefont {Song}, \citenamefont {Kivshar},\ and\ \citenamefont {Qiu}}]{RN74}%
  \BibitemOpen
  \bibfield  {author} {\bibinfo {author} {\bibfnamefont {J.}~\bibnamefont {Ni}}, \bibinfo {author} {\bibfnamefont {C.}~\bibnamefont {Huang}}, \bibinfo {author} {\bibfnamefont {L.-M.}\ \bibnamefont {Zhou}}, \bibinfo {author} {\bibfnamefont {M.}~\bibnamefont {Gu}}, \bibinfo {author} {\bibfnamefont {Q.}~\bibnamefont {Song}}, \bibinfo {author} {\bibfnamefont {Y.}~\bibnamefont {Kivshar}},\ and\ \bibinfo {author} {\bibfnamefont {C.-W.}\ \bibnamefont {Qiu}},\ }\bibfield  {title} {\bibinfo {title} {Multidimensional phase singularities in nanophotonics},\ }\href {https://doi.org/10.1126/science.abj0039} {\bibfield  {journal} {\bibinfo  {journal} {Science}\ }\textbf {\bibinfo {volume} {374}},\ \bibinfo {pages} {eabj0039} (\bibinfo {year} {2021})}\BibitemShut {NoStop}%
\bibitem [{\citenamefont {Wang}\ \emph {et~al.}(2015)\citenamefont {Wang}, \citenamefont {Cai}, \citenamefont {Su}, \citenamefont {Chen}, \citenamefont {Wu}, \citenamefont {Li}, \citenamefont {Liu}, \citenamefont {Lu},\ and\ \citenamefont {Pan}}]{RN11}%
  \BibitemOpen
  \bibfield  {author} {\bibinfo {author} {\bibfnamefont {X.-L.}\ \bibnamefont {Wang}}, \bibinfo {author} {\bibfnamefont {X.-D.}\ \bibnamefont {Cai}}, \bibinfo {author} {\bibfnamefont {Z.-E.}\ \bibnamefont {Su}}, \bibinfo {author} {\bibfnamefont {M.-C.}\ \bibnamefont {Chen}}, \bibinfo {author} {\bibfnamefont {D.}~\bibnamefont {Wu}}, \bibinfo {author} {\bibfnamefont {L.}~\bibnamefont {Li}}, \bibinfo {author} {\bibfnamefont {N.-L.}\ \bibnamefont {Liu}}, \bibinfo {author} {\bibfnamefont {C.-Y.}\ \bibnamefont {Lu}},\ and\ \bibinfo {author} {\bibfnamefont {J.-W.}\ \bibnamefont {Pan}},\ }\bibfield  {title} {\bibinfo {title} {Quantum teleportation of multiple degrees of freedom of a single photon},\ }\href@noop {} {\bibfield  {journal} {\bibinfo  {journal} {Nature}\ }\textbf {\bibinfo {volume} {518}},\ \bibinfo {pages} {516} (\bibinfo {year} {2015})}\BibitemShut {NoStop}%
\bibitem [{\citenamefont {Shen}\ \emph {et~al.}(2019)\citenamefont {Shen}, \citenamefont {Wang}, \citenamefont {Xie}, \citenamefont {Min}, \citenamefont {Fu}, \citenamefont {Liu}, \citenamefont {Gong},\ and\ \citenamefont {Yuan}}]{RN70}%
  \BibitemOpen
  \bibfield  {author} {\bibinfo {author} {\bibfnamefont {Y.}~\bibnamefont {Shen}}, \bibinfo {author} {\bibfnamefont {X.}~\bibnamefont {Wang}}, \bibinfo {author} {\bibfnamefont {Z.}~\bibnamefont {Xie}}, \bibinfo {author} {\bibfnamefont {C.}~\bibnamefont {Min}}, \bibinfo {author} {\bibfnamefont {X.}~\bibnamefont {Fu}}, \bibinfo {author} {\bibfnamefont {Q.}~\bibnamefont {Liu}}, \bibinfo {author} {\bibfnamefont {M.}~\bibnamefont {Gong}},\ and\ \bibinfo {author} {\bibfnamefont {X.}~\bibnamefont {Yuan}},\ }\bibfield  {title} {\bibinfo {title} {Optical vortices 30 years on: Oam manipulation from topological charge to multiple singularities},\ }\href@noop {} {\bibfield  {journal} {\bibinfo  {journal} {Light Sci. Appl.}\ }\textbf {\bibinfo {volume} {8}},\ \bibinfo {pages} {90} (\bibinfo {year} {2019})}\BibitemShut {NoStop}%
\bibitem [{\citenamefont {Bozinovic}\ \emph {et~al.}(2013)\citenamefont {Bozinovic}, \citenamefont {Yue}, \citenamefont {Ren}, \citenamefont {Tur}, \citenamefont {Kristensen}, \citenamefont {Huang}, \citenamefont {Willner},\ and\ \citenamefont {Ramachandran}}]{RN14}%
  \BibitemOpen
  \bibfield  {author} {\bibinfo {author} {\bibfnamefont {N.}~\bibnamefont {Bozinovic}}, \bibinfo {author} {\bibfnamefont {Y.}~\bibnamefont {Yue}}, \bibinfo {author} {\bibfnamefont {Y.}~\bibnamefont {Ren}}, \bibinfo {author} {\bibfnamefont {M.}~\bibnamefont {Tur}}, \bibinfo {author} {\bibfnamefont {P.}~\bibnamefont {Kristensen}}, \bibinfo {author} {\bibfnamefont {H.}~\bibnamefont {Huang}}, \bibinfo {author} {\bibfnamefont {A.~E.}\ \bibnamefont {Willner}},\ and\ \bibinfo {author} {\bibfnamefont {S.}~\bibnamefont {Ramachandran}},\ }\bibfield  {title} {\bibinfo {title} {Terabit-scale orbital angular momentum mode division multiplexing in fibers},\ }\href@noop {} {\bibfield  {journal} {\bibinfo  {journal} {Science}\ }\textbf {\bibinfo {volume} {340}},\ \bibinfo {pages} {1545} (\bibinfo {year} {2013})}\BibitemShut {NoStop}%
\bibitem [{\citenamefont {Liu}\ \emph {et~al.}(2020)\citenamefont {Liu}, \citenamefont {Lou},\ and\ \citenamefont {Jing}}]{RN12}%
  \BibitemOpen
  \bibfield  {author} {\bibinfo {author} {\bibfnamefont {S.}~\bibnamefont {Liu}}, \bibinfo {author} {\bibfnamefont {Y.}~\bibnamefont {Lou}},\ and\ \bibinfo {author} {\bibfnamefont {J.}~\bibnamefont {Jing}},\ }\bibfield  {title} {\bibinfo {title} {Orbital angular momentum multiplexed deterministic all-optical quantum teleportation},\ }\href@noop {} {\bibfield  {journal} {\bibinfo  {journal} {Nat. Commun.}\ }\textbf {\bibinfo {volume} {11}},\ \bibinfo {pages} {3875} (\bibinfo {year} {2020})}\BibitemShut {NoStop}%
\bibitem [{\citenamefont {Chen}\ \emph {et~al.}(2021{\natexlab{a}})\citenamefont {Chen}, \citenamefont {Yang}, \citenamefont {Chen},\ and\ \citenamefont {Liu}}]{RN47}%
  \BibitemOpen
  \bibfield  {author} {\bibinfo {author} {\bibfnamefont {W.}~\bibnamefont {Chen}}, \bibinfo {author} {\bibfnamefont {Q.}~\bibnamefont {Yang}}, \bibinfo {author} {\bibfnamefont {Y.}~\bibnamefont {Chen}},\ and\ \bibinfo {author} {\bibfnamefont {W.}~\bibnamefont {Liu}},\ }\bibfield  {title} {\bibinfo {title} {Evolution and global charge conservation for polarization singularities emerging from non-hermitian degeneracies},\ }\href@noop {} {\bibfield  {journal} {\bibinfo  {journal} {Proc. Natl. Acad. Sci. U. S. A.}\ }\textbf {\bibinfo {volume} {118}},\ \bibinfo {pages} {e2019578118} (\bibinfo {year} {2021}{\natexlab{a}})}\BibitemShut {NoStop}%
\bibitem [{\citenamefont {Spaegele}\ \emph {et~al.}(2023)\citenamefont {Spaegele}, \citenamefont {Tamagnone}, \citenamefont {Lim}, \citenamefont {Ossiander}, \citenamefont {Meretska},\ and\ \citenamefont {Capasso}}]{RN61}%
  \BibitemOpen
  \bibfield  {author} {\bibinfo {author} {\bibfnamefont {C.~M.}\ \bibnamefont {Spaegele}}, \bibinfo {author} {\bibfnamefont {M.}~\bibnamefont {Tamagnone}}, \bibinfo {author} {\bibfnamefont {S.~W.~D.}\ \bibnamefont {Lim}}, \bibinfo {author} {\bibfnamefont {M.}~\bibnamefont {Ossiander}}, \bibinfo {author} {\bibfnamefont {M.~L.}\ \bibnamefont {Meretska}},\ and\ \bibinfo {author} {\bibfnamefont {F.}~\bibnamefont {Capasso}},\ }\bibfield  {title} {\bibinfo {title} {Topologically protected optical polarization singularities in four-dimensional space},\ }\href@noop {} {\bibfield  {journal} {\bibinfo  {journal} {Sci. Adv.}\ }\textbf {\bibinfo {volume} {9}},\ \bibinfo {pages} {eadh0369} (\bibinfo {year} {2023})}\BibitemShut {NoStop}%
\bibitem [{\citenamefont {Chen}\ \emph {et~al.}(2023)\citenamefont {Chen}, \citenamefont {Deng}, \citenamefont {Sha}, \citenamefont {Chen}, \citenamefont {Wang}, \citenamefont {Chen}, \citenamefont {Wu}, \citenamefont {Chu}, \citenamefont {Kivshar}, \citenamefont {Xiao},\ and\ \citenamefont {Qiu}}]{RN73}%
  \BibitemOpen
  \bibfield  {author} {\bibinfo {author} {\bibfnamefont {Y.}~\bibnamefont {Chen}}, \bibinfo {author} {\bibfnamefont {H.}~\bibnamefont {Deng}}, \bibinfo {author} {\bibfnamefont {X.}~\bibnamefont {Sha}}, \bibinfo {author} {\bibfnamefont {W.}~\bibnamefont {Chen}}, \bibinfo {author} {\bibfnamefont {R.}~\bibnamefont {Wang}}, \bibinfo {author} {\bibfnamefont {Y.-H.}\ \bibnamefont {Chen}}, \bibinfo {author} {\bibfnamefont {D.}~\bibnamefont {Wu}}, \bibinfo {author} {\bibfnamefont {J.}~\bibnamefont {Chu}}, \bibinfo {author} {\bibfnamefont {Y.~S.}\ \bibnamefont {Kivshar}}, \bibinfo {author} {\bibfnamefont {S.}~\bibnamefont {Xiao}},\ and\ \bibinfo {author} {\bibfnamefont {C.-W.}\ \bibnamefont {Qiu}},\ }\bibfield  {title} {\bibinfo {title} {Observation of intrinsic chiral bound states in the continuum},\ }\href@noop {} {\bibfield  {journal} {\bibinfo  {journal} {Nature}\ }\textbf {\bibinfo {volume} {613}},\ \bibinfo {pages} {474} (\bibinfo {year} {2023})}\BibitemShut {NoStop}%
\bibitem [{\citenamefont {Ye}\ \emph {et~al.}(2020)\citenamefont {Ye}, \citenamefont {Gao},\ and\ \citenamefont {Liu}}]{RN55}%
  \BibitemOpen
  \bibfield  {author} {\bibinfo {author} {\bibfnamefont {W.}~\bibnamefont {Ye}}, \bibinfo {author} {\bibfnamefont {Y.}~\bibnamefont {Gao}},\ and\ \bibinfo {author} {\bibfnamefont {J.}~\bibnamefont {Liu}},\ }\bibfield  {title} {\bibinfo {title} {Singular points of polarizations in the momentum space of photonic crystal slabs},\ }\href@noop {} {\bibfield  {journal} {\bibinfo  {journal} {Phys. Rev. Lett.}\ }\textbf {\bibinfo {volume} {124}},\ \bibinfo {pages} {153904} (\bibinfo {year} {2020})}\BibitemShut {NoStop}%
\bibitem [{\citenamefont {Flossmann}\ \emph {et~al.}(2005)\citenamefont {Flossmann}, \citenamefont {Schwarz}, \citenamefont {Maier},\ and\ \citenamefont {Dennis}}]{RN68}%
  \BibitemOpen
  \bibfield  {author} {\bibinfo {author} {\bibfnamefont {F.}~\bibnamefont {Flossmann}}, \bibinfo {author} {\bibfnamefont {U.~T.}\ \bibnamefont {Schwarz}}, \bibinfo {author} {\bibfnamefont {M.}~\bibnamefont {Maier}},\ and\ \bibinfo {author} {\bibfnamefont {M.~R.}\ \bibnamefont {Dennis}},\ }\bibfield  {title} {\bibinfo {title} {Polarization singularities from unfolding an optical vortex through a birefringent crystal},\ }\href@noop {} {\bibfield  {journal} {\bibinfo  {journal} {Phys. Rev. Lett.}\ }\textbf {\bibinfo {volume} {95}},\ \bibinfo {pages} {253901} (\bibinfo {year} {2005})}\BibitemShut {NoStop}%
\bibitem [{\citenamefont {Liu}\ \emph {et~al.}(2019)\citenamefont {Liu}, \citenamefont {Wang}, \citenamefont {Zhang}, \citenamefont {Wang}, \citenamefont {Zhao}, \citenamefont {Guan}, \citenamefont {Liu}, \citenamefont {Shi},\ and\ \citenamefont {Zi}}]{RN63}%
  \BibitemOpen
  \bibfield  {author} {\bibinfo {author} {\bibfnamefont {W.}~\bibnamefont {Liu}}, \bibinfo {author} {\bibfnamefont {B.}~\bibnamefont {Wang}}, \bibinfo {author} {\bibfnamefont {Y.}~\bibnamefont {Zhang}}, \bibinfo {author} {\bibfnamefont {J.}~\bibnamefont {Wang}}, \bibinfo {author} {\bibfnamefont {M.}~\bibnamefont {Zhao}}, \bibinfo {author} {\bibfnamefont {F.}~\bibnamefont {Guan}}, \bibinfo {author} {\bibfnamefont {X.}~\bibnamefont {Liu}}, \bibinfo {author} {\bibfnamefont {L.}~\bibnamefont {Shi}},\ and\ \bibinfo {author} {\bibfnamefont {J.}~\bibnamefont {Zi}},\ }\bibfield  {title} {\bibinfo {title} {Circularly polarized states spawning from bound states in the continuum},\ }\href@noop {} {\bibfield  {journal} {\bibinfo  {journal} {Phys. Rev. Lett.}\ }\textbf {\bibinfo {volume} {123}},\ \bibinfo {pages} {116104} (\bibinfo {year} {2019})}\BibitemShut {NoStop}%
\bibitem [{\citenamefont {Lim}\ \emph {et~al.}(2021)\citenamefont {Lim}, \citenamefont {Park}, \citenamefont {Meretska}, \citenamefont {Dorrah},\ and\ \citenamefont {Capasso}}]{RN64}%
  \BibitemOpen
  \bibfield  {author} {\bibinfo {author} {\bibfnamefont {S.~W.~D.}\ \bibnamefont {Lim}}, \bibinfo {author} {\bibfnamefont {J.-S.}\ \bibnamefont {Park}}, \bibinfo {author} {\bibfnamefont {M.~L.}\ \bibnamefont {Meretska}}, \bibinfo {author} {\bibfnamefont {A.~H.}\ \bibnamefont {Dorrah}},\ and\ \bibinfo {author} {\bibfnamefont {F.}~\bibnamefont {Capasso}},\ }\bibfield  {title} {\bibinfo {title} {Engineering phase and polarization singularity sheets},\ }\href@noop {} {\bibfield  {journal} {\bibinfo  {journal} {Nat. Commun.}\ }\textbf {\bibinfo {volume} {12}},\ \bibinfo {pages} {4190} (\bibinfo {year} {2021})}\BibitemShut {NoStop}%
\bibitem [{\citenamefont {Overvig}\ \emph {et~al.}(2021)\citenamefont {Overvig}, \citenamefont {Yu},\ and\ \citenamefont {Alù}}]{RN52}%
  \BibitemOpen
  \bibfield  {author} {\bibinfo {author} {\bibfnamefont {A.}~\bibnamefont {Overvig}}, \bibinfo {author} {\bibfnamefont {N.}~\bibnamefont {Yu}},\ and\ \bibinfo {author} {\bibfnamefont {A.}~\bibnamefont {Alù}},\ }\bibfield  {title} {\bibinfo {title} {Chiral quasi-bound states in the continuum},\ }\href@noop {} {\bibfield  {journal} {\bibinfo  {journal} {Phys. Rev. Lett.}\ }\textbf {\bibinfo {volume} {126}} (\bibinfo {year} {2021})}\BibitemShut {NoStop}%
\bibitem [{\citenamefont {Yoda}\ and\ \citenamefont {Notomi}(2020)}]{RN58}%
  \BibitemOpen
  \bibfield  {author} {\bibinfo {author} {\bibfnamefont {T.}~\bibnamefont {Yoda}}\ and\ \bibinfo {author} {\bibfnamefont {M.}~\bibnamefont {Notomi}},\ }\bibfield  {title} {\bibinfo {title} {Generation and annihilation of topologically protected bound states in the continuum and circularly polarized states by symmetry breaking},\ }\href@noop {} {\bibfield  {journal} {\bibinfo  {journal} {Phys. Rev. Lett.}\ }\textbf {\bibinfo {volume} {125}},\ \bibinfo {pages} {053902} (\bibinfo {year} {2020})}\BibitemShut {NoStop}%
\bibitem [{\citenamefont {Jin}\ \emph {et~al.}(2019)\citenamefont {Jin}, \citenamefont {Yin}, \citenamefont {Ni}, \citenamefont {Soljačić}, \citenamefont {Zhen},\ and\ \citenamefont {Peng}}]{RN30}%
  \BibitemOpen
  \bibfield  {author} {\bibinfo {author} {\bibfnamefont {J.}~\bibnamefont {Jin}}, \bibinfo {author} {\bibfnamefont {X.}~\bibnamefont {Yin}}, \bibinfo {author} {\bibfnamefont {L.}~\bibnamefont {Ni}}, \bibinfo {author} {\bibfnamefont {M.}~\bibnamefont {Soljačić}}, \bibinfo {author} {\bibfnamefont {B.}~\bibnamefont {Zhen}},\ and\ \bibinfo {author} {\bibfnamefont {C.}~\bibnamefont {Peng}},\ }\bibfield  {title} {\bibinfo {title} {Topologically enabled ultrahigh-q guided resonances robust to out-of-plane scattering},\ }\href@noop {} {\bibfield  {journal} {\bibinfo  {journal} {Nature}\ }\textbf {\bibinfo {volume} {574}},\ \bibinfo {pages} {501} (\bibinfo {year} {2019})}\BibitemShut {NoStop}%
\bibitem [{\citenamefont {Chen}\ \emph {et~al.}(2022{\natexlab{a}})\citenamefont {Chen}, \citenamefont {Yin}, \citenamefont {Jin}, \citenamefont {Zheng}, \citenamefont {Zhang}, \citenamefont {Wang}, \citenamefont {He}, \citenamefont {Zhen},\ and\ \citenamefont {Peng}}]{RN72}%
  \BibitemOpen
  \bibfield  {author} {\bibinfo {author} {\bibfnamefont {Z.}~\bibnamefont {Chen}}, \bibinfo {author} {\bibfnamefont {X.}~\bibnamefont {Yin}}, \bibinfo {author} {\bibfnamefont {J.}~\bibnamefont {Jin}}, \bibinfo {author} {\bibfnamefont {Z.}~\bibnamefont {Zheng}}, \bibinfo {author} {\bibfnamefont {Z.}~\bibnamefont {Zhang}}, \bibinfo {author} {\bibfnamefont {F.}~\bibnamefont {Wang}}, \bibinfo {author} {\bibfnamefont {L.}~\bibnamefont {He}}, \bibinfo {author} {\bibfnamefont {B.}~\bibnamefont {Zhen}},\ and\ \bibinfo {author} {\bibfnamefont {C.}~\bibnamefont {Peng}},\ }\bibfield  {title} {\bibinfo {title} {Observation of miniaturized bound states in the continuum with ultra-high quality factors},\ }\href@noop {} {\bibfield  {journal} {\bibinfo  {journal} {Sci. Bull. (Beijing)}\ }\textbf {\bibinfo {volume} {67}},\ \bibinfo {pages} {359} (\bibinfo {year} {2022}{\natexlab{a}})}\BibitemShut {NoStop}%
\bibitem [{\citenamefont {Yang}\ \emph {et~al.}(2014)\citenamefont {Yang}, \citenamefont {Peng}, \citenamefont {Liang}, \citenamefont {Li},\ and\ \citenamefont {Noda}}]{RN51}%
  \BibitemOpen
  \bibfield  {author} {\bibinfo {author} {\bibfnamefont {Y.}~\bibnamefont {Yang}}, \bibinfo {author} {\bibfnamefont {C.}~\bibnamefont {Peng}}, \bibinfo {author} {\bibfnamefont {Y.}~\bibnamefont {Liang}}, \bibinfo {author} {\bibfnamefont {Z.}~\bibnamefont {Li}},\ and\ \bibinfo {author} {\bibfnamefont {S.}~\bibnamefont {Noda}},\ }\bibfield  {title} {\bibinfo {title} {Analytical perspective for bound states in the continuum in photonic crystal slabs},\ }\href@noop {} {\bibfield  {journal} {\bibinfo  {journal} {Phys. Rev. Lett.}\ }\textbf {\bibinfo {volume} {113}},\ \bibinfo {pages} {037401} (\bibinfo {year} {2014})}\BibitemShut {NoStop}%
\bibitem [{\citenamefont {Jiang}\ \emph {et~al.}(2023)\citenamefont {Jiang}, \citenamefont {Hu}, \citenamefont {Wang}, \citenamefont {Han},\ and\ \citenamefont {Zi}}]{RN54}%
  \BibitemOpen
  \bibfield  {author} {\bibinfo {author} {\bibfnamefont {Q.}~\bibnamefont {Jiang}}, \bibinfo {author} {\bibfnamefont {P.}~\bibnamefont {Hu}}, \bibinfo {author} {\bibfnamefont {J.}~\bibnamefont {Wang}}, \bibinfo {author} {\bibfnamefont {D.}~\bibnamefont {Han}},\ and\ \bibinfo {author} {\bibfnamefont {J.}~\bibnamefont {Zi}},\ }\bibfield  {title} {\bibinfo {title} {General bound states in the continuum in momentum space},\ }\href@noop {} {\bibfield  {journal} {\bibinfo  {journal} {Phys. Rev. Lett.}\ }\textbf {\bibinfo {volume} {131}},\ \bibinfo {pages} {013801} (\bibinfo {year} {2023})}\BibitemShut {NoStop}%
\bibitem [{\citenamefont {Kang}\ \emph {et~al.}(2021)\citenamefont {Kang}, \citenamefont {Zhang}, \citenamefont {Xiao},\ and\ \citenamefont {Xu}}]{RN53}%
  \BibitemOpen
  \bibfield  {author} {\bibinfo {author} {\bibfnamefont {M.}~\bibnamefont {Kang}}, \bibinfo {author} {\bibfnamefont {S.}~\bibnamefont {Zhang}}, \bibinfo {author} {\bibfnamefont {M.}~\bibnamefont {Xiao}},\ and\ \bibinfo {author} {\bibfnamefont {H.}~\bibnamefont {Xu}},\ }\bibfield  {title} {\bibinfo {title} {Merging bound states in the continuum at off-high symmetry points},\ }\href@noop {} {\bibfield  {journal} {\bibinfo  {journal} {Phys. Rev. Lett.}\ }\textbf {\bibinfo {volume} {126}},\ \bibinfo {pages} {117402} (\bibinfo {year} {2021})}\BibitemShut {NoStop}%
\bibitem [{\citenamefont {Hu}\ \emph {et~al.}(2022)\citenamefont {Hu}, \citenamefont {Wang}, \citenamefont {Jiang}, \citenamefont {Wang}, \citenamefont {Shi}, \citenamefont {Han}, \citenamefont {Zhang}, \citenamefont {Chan},\ and\ \citenamefont {Zi}}]{RN49}%
  \BibitemOpen
  \bibfield  {author} {\bibinfo {author} {\bibfnamefont {P.}~\bibnamefont {Hu}}, \bibinfo {author} {\bibfnamefont {J.}~\bibnamefont {Wang}}, \bibinfo {author} {\bibfnamefont {Q.}~\bibnamefont {Jiang}}, \bibinfo {author} {\bibfnamefont {J.}~\bibnamefont {Wang}}, \bibinfo {author} {\bibfnamefont {L.}~\bibnamefont {Shi}}, \bibinfo {author} {\bibfnamefont {D.}~\bibnamefont {Han}}, \bibinfo {author} {\bibfnamefont {Z.~Q.}\ \bibnamefont {Zhang}}, \bibinfo {author} {\bibfnamefont {C.~T.}\ \bibnamefont {Chan}},\ and\ \bibinfo {author} {\bibfnamefont {J.}~\bibnamefont {Zi}},\ }\bibfield  {title} {\bibinfo {title} {Global phase diagram of bound states in the continuum},\ }\href@noop {} {\bibfield  {journal} {\bibinfo  {journal} {Optica}\ }\textbf {\bibinfo {volume} {9}},\ \bibinfo {pages} {1353} (\bibinfo {year} {2022})}\BibitemShut {NoStop}%
\bibitem [{\citenamefont {Yin}\ \emph {et~al.}(2020)\citenamefont {Yin}, \citenamefont {Jin}, \citenamefont {Soljačić}, \citenamefont {Peng},\ and\ \citenamefont {Zhen}}]{RN27}%
  \BibitemOpen
  \bibfield  {author} {\bibinfo {author} {\bibfnamefont {X.}~\bibnamefont {Yin}}, \bibinfo {author} {\bibfnamefont {J.}~\bibnamefont {Jin}}, \bibinfo {author} {\bibfnamefont {M.}~\bibnamefont {Soljačić}}, \bibinfo {author} {\bibfnamefont {C.}~\bibnamefont {Peng}},\ and\ \bibinfo {author} {\bibfnamefont {B.}~\bibnamefont {Zhen}},\ }\bibfield  {title} {\bibinfo {title} {Observation of topologically enabled unidirectional guided resonances},\ }\href@noop {} {\bibfield  {journal} {\bibinfo  {journal} {Nature}\ }\textbf {\bibinfo {volume} {580}},\ \bibinfo {pages} {467} (\bibinfo {year} {2020})}\BibitemShut {NoStop}%
\bibitem [{\citenamefont {Yin}\ \emph {et~al.}(2023)\citenamefont {Yin}, \citenamefont {Inoue}, \citenamefont {Peng},\ and\ \citenamefont {Noda}}]{RN4}%
  \BibitemOpen
  \bibfield  {author} {\bibinfo {author} {\bibfnamefont {X.}~\bibnamefont {Yin}}, \bibinfo {author} {\bibfnamefont {T.}~\bibnamefont {Inoue}}, \bibinfo {author} {\bibfnamefont {C.}~\bibnamefont {Peng}},\ and\ \bibinfo {author} {\bibfnamefont {S.}~\bibnamefont {Noda}},\ }\bibfield  {title} {\bibinfo {title} {Topological unidirectional guided resonances emerged from interband coupling},\ }\href@noop {} {\bibfield  {journal} {\bibinfo  {journal} {Phys. Rev. Lett.}\ }\textbf {\bibinfo {volume} {130}},\ \bibinfo {pages} {056401} (\bibinfo {year} {2023})}\BibitemShut {NoStop}%
\bibitem [{\citenamefont {Gong}\ \emph {et~al.}(2023)\citenamefont {Gong}, \citenamefont {Liu}, \citenamefont {Ge}, \citenamefont {Xiang},\ and\ \citenamefont {Han}}]{RN56}%
  \BibitemOpen
  \bibfield  {author} {\bibinfo {author} {\bibfnamefont {M.}~\bibnamefont {Gong}}, \bibinfo {author} {\bibfnamefont {J.}~\bibnamefont {Liu}}, \bibinfo {author} {\bibfnamefont {L.}~\bibnamefont {Ge}}, \bibinfo {author} {\bibfnamefont {H.}~\bibnamefont {Xiang}},\ and\ \bibinfo {author} {\bibfnamefont {D.}~\bibnamefont {Han}},\ }\bibfield  {title} {\bibinfo {title} {Multipolar perspective on unidirectional guided resonances},\ }\href@noop {} {\bibfield  {journal} {\bibinfo  {journal} {Phys. Rev. A}\ }\textbf {\bibinfo {volume} {108}} (\bibinfo {year} {2023})}\BibitemShut {NoStop}%
\bibitem [{\citenamefont {Allen}\ \emph {et~al.}(1992)\citenamefont {Allen}, \citenamefont {Beijersbergen}, \citenamefont {Spreeuw},\ and\ \citenamefont {Woerdman}}]{RN50}%
  \BibitemOpen
  \bibfield  {author} {\bibinfo {author} {\bibfnamefont {L.}~\bibnamefont {Allen}}, \bibinfo {author} {\bibfnamefont {M.~W.}\ \bibnamefont {Beijersbergen}}, \bibinfo {author} {\bibfnamefont {R.~J.}\ \bibnamefont {Spreeuw}},\ and\ \bibinfo {author} {\bibfnamefont {J.~P.}\ \bibnamefont {Woerdman}},\ }\bibfield  {title} {\bibinfo {title} {Orbital angular momentum of light and the transformation of laguerre-gaussian laser modes},\ }\href@noop {} {\bibfield  {journal} {\bibinfo  {journal} {Phys. Rev. A}\ }\textbf {\bibinfo {volume} {45}},\ \bibinfo {pages} {8185} (\bibinfo {year} {1992})}\BibitemShut {NoStop}%
\bibitem [{\citenamefont {Dorrah}\ \emph {et~al.}(2018)\citenamefont {Dorrah}, \citenamefont {Rosales-Guzmán}, \citenamefont {Forbes},\ and\ \citenamefont {Mojahedi}}]{RN69}%
  \BibitemOpen
  \bibfield  {author} {\bibinfo {author} {\bibfnamefont {A.~H.}\ \bibnamefont {Dorrah}}, \bibinfo {author} {\bibfnamefont {C.}~\bibnamefont {Rosales-Guzmán}}, \bibinfo {author} {\bibfnamefont {A.}~\bibnamefont {Forbes}},\ and\ \bibinfo {author} {\bibfnamefont {M.}~\bibnamefont {Mojahedi}},\ }\bibfield  {title} {\bibinfo {title} {Evolution of orbital angular momentum in three-dimensional structured light},\ }\href@noop {} {\bibfield  {journal} {\bibinfo  {journal} {Phys. Rev. A}\ }\textbf {\bibinfo {volume} {98}} (\bibinfo {year} {2018})}\BibitemShut {NoStop}%
\bibitem [{\citenamefont {Qiao}\ \emph {et~al.}(2023)\citenamefont {Qiao}, \citenamefont {Zhu}, \citenamefont {Yuan}, \citenamefont {Gong}, \citenamefont {Liao}, \citenamefont {Gong}, \citenamefont {Kim},\ and\ \citenamefont {Chen}}]{RN48}%
  \BibitemOpen
  \bibfield  {author} {\bibinfo {author} {\bibfnamefont {Z.}~\bibnamefont {Qiao}}, \bibinfo {author} {\bibfnamefont {S.}~\bibnamefont {Zhu}}, \bibinfo {author} {\bibfnamefont {Z.}~\bibnamefont {Yuan}}, \bibinfo {author} {\bibfnamefont {C.}~\bibnamefont {Gong}}, \bibinfo {author} {\bibfnamefont {Y.}~\bibnamefont {Liao}}, \bibinfo {author} {\bibfnamefont {X.}~\bibnamefont {Gong}}, \bibinfo {author} {\bibfnamefont {M.}~\bibnamefont {Kim}},\ and\ \bibinfo {author} {\bibfnamefont {Y.-C.}\ \bibnamefont {Chen}},\ }\bibfield  {title} {\bibinfo {title} {High orbital angular momentum lasing with tunable degree of chirality in a symmetry-broken microcavity}} (\bibinfo {year} {2023})\BibitemShut {NoStop}%
\bibitem [{\citenamefont {Chen}\ \emph {et~al.}(2022{\natexlab{b}})\citenamefont {Chen}, \citenamefont {Zhu}, \citenamefont {Huo}, \citenamefont {Song}, \citenamefont {Lezec}, \citenamefont {Xu},\ and\ \citenamefont {Agrawal}}]{RN8}%
  \BibitemOpen
  \bibfield  {author} {\bibinfo {author} {\bibfnamefont {L.}~\bibnamefont {Chen}}, \bibinfo {author} {\bibfnamefont {W.}~\bibnamefont {Zhu}}, \bibinfo {author} {\bibfnamefont {P.}~\bibnamefont {Huo}}, \bibinfo {author} {\bibfnamefont {J.}~\bibnamefont {Song}}, \bibinfo {author} {\bibfnamefont {H.~J.}\ \bibnamefont {Lezec}}, \bibinfo {author} {\bibfnamefont {T.}~\bibnamefont {Xu}},\ and\ \bibinfo {author} {\bibfnamefont {A.}~\bibnamefont {Agrawal}},\ }\bibfield  {title} {\bibinfo {title} {Synthesizing ultrafast optical pulses with arbitrary spatiotemporal control},\ }\href@noop {} {\bibfield  {journal} {\bibinfo  {journal} {Sci. Adv.}\ }\textbf {\bibinfo {volume} {8}},\ \bibinfo {pages} {eabq8314} (\bibinfo {year} {2022}{\natexlab{b}})}\BibitemShut {NoStop}%
\bibitem [{\citenamefont {Chong}\ \emph {et~al.}(2020)\citenamefont {Chong}, \citenamefont {Wan}, \citenamefont {Chen},\ and\ \citenamefont {Zhan}}]{RN66}%
  \BibitemOpen
  \bibfield  {author} {\bibinfo {author} {\bibfnamefont {A.}~\bibnamefont {Chong}}, \bibinfo {author} {\bibfnamefont {C.}~\bibnamefont {Wan}}, \bibinfo {author} {\bibfnamefont {J.}~\bibnamefont {Chen}},\ and\ \bibinfo {author} {\bibfnamefont {Q.}~\bibnamefont {Zhan}},\ }\bibfield  {title} {\bibinfo {title} {Generation of spatiotemporal optical vortices with controllable transverse orbital angular momentum},\ }\href@noop {} {\bibfield  {journal} {\bibinfo  {journal} {Nat. Photonics}\ }\textbf {\bibinfo {volume} {14}},\ \bibinfo {pages} {350} (\bibinfo {year} {2020})}\BibitemShut {NoStop}%
\bibitem [{\citenamefont {Huang}\ \emph {et~al.}(2022)\citenamefont {Huang}, \citenamefont {Zhang}, \citenamefont {Zhu},\ and\ \citenamefont {Ruan}}]{RN65}%
  \BibitemOpen
  \bibfield  {author} {\bibinfo {author} {\bibfnamefont {J.}~\bibnamefont {Huang}}, \bibinfo {author} {\bibfnamefont {J.}~\bibnamefont {Zhang}}, \bibinfo {author} {\bibfnamefont {T.}~\bibnamefont {Zhu}},\ and\ \bibinfo {author} {\bibfnamefont {Z.}~\bibnamefont {Ruan}},\ }\bibfield  {title} {\bibinfo {title} {Spatiotemporal differentiators generating optical vortices with transverse orbital angular momentum and detecting sharp change of pulse envelope},\ }\href@noop {} {\bibfield  {journal} {\bibinfo  {journal} {Laser Photon. Rev.}\ }\textbf {\bibinfo {volume} {16}},\ \bibinfo {pages} {2100357} (\bibinfo {year} {2022})}\BibitemShut {NoStop}%
\bibitem [{\citenamefont {Yang}\ \emph {et~al.}(2020)\citenamefont {Yang}, \citenamefont {Shao}, \citenamefont {Chen}, \citenamefont {Mao},\ and\ \citenamefont {Ma}}]{RN38}%
  \BibitemOpen
  \bibfield  {author} {\bibinfo {author} {\bibfnamefont {Z.-Q.}\ \bibnamefont {Yang}}, \bibinfo {author} {\bibfnamefont {Z.-K.}\ \bibnamefont {Shao}}, \bibinfo {author} {\bibfnamefont {H.-Z.}\ \bibnamefont {Chen}}, \bibinfo {author} {\bibfnamefont {X.-R.}\ \bibnamefont {Mao}},\ and\ \bibinfo {author} {\bibfnamefont {R.-M.}\ \bibnamefont {Ma}},\ }\bibfield  {title} {\bibinfo {title} {Spin-momentum-locked edge mode for topological vortex lasing},\ }\href@noop {} {\bibfield  {journal} {\bibinfo  {journal} {Phys. Rev. Lett.}\ }\textbf {\bibinfo {volume} {125}},\ \bibinfo {pages} {013903} (\bibinfo {year} {2020})}\BibitemShut {NoStop}%
\bibitem [{\citenamefont {Sroor}\ \emph {et~al.}(2020)\citenamefont {Sroor}, \citenamefont {Huang}, \citenamefont {Sephton}, \citenamefont {Naidoo}, \citenamefont {Vallés}, \citenamefont {Ginis}, \citenamefont {Qiu}, \citenamefont {Ambrosio}, \citenamefont {Capasso},\ and\ \citenamefont {Forbes}}]{RN40}%
  \BibitemOpen
  \bibfield  {author} {\bibinfo {author} {\bibfnamefont {H.}~\bibnamefont {Sroor}}, \bibinfo {author} {\bibfnamefont {Y.-W.}\ \bibnamefont {Huang}}, \bibinfo {author} {\bibfnamefont {B.}~\bibnamefont {Sephton}}, \bibinfo {author} {\bibfnamefont {D.}~\bibnamefont {Naidoo}}, \bibinfo {author} {\bibfnamefont {A.}~\bibnamefont {Vallés}}, \bibinfo {author} {\bibfnamefont {V.}~\bibnamefont {Ginis}}, \bibinfo {author} {\bibfnamefont {C.-W.}\ \bibnamefont {Qiu}}, \bibinfo {author} {\bibfnamefont {A.}~\bibnamefont {Ambrosio}}, \bibinfo {author} {\bibfnamefont {F.}~\bibnamefont {Capasso}},\ and\ \bibinfo {author} {\bibfnamefont {A.}~\bibnamefont {Forbes}},\ }\bibfield  {title} {\bibinfo {title} {High-purity orbital angular momentum states from a visible metasurface laser},\ }\href@noop {} {\bibfield  {journal} {\bibinfo  {journal} {Nat. Photonics}\ }\textbf {\bibinfo {volume} {14}},\ \bibinfo {pages} {498} (\bibinfo {year} {2020})}\BibitemShut {NoStop}%
\bibitem [{\citenamefont {Raun}\ \emph {et~al.}(2023)\citenamefont {Raun}, \citenamefont {Tang}, \citenamefont {Ni}, \citenamefont {Mazur},\ and\ \citenamefont {Hu}}]{RN77}%
  \BibitemOpen
  \bibfield  {author} {\bibinfo {author} {\bibfnamefont {A.}~\bibnamefont {Raun}}, \bibinfo {author} {\bibfnamefont {H.}~\bibnamefont {Tang}}, \bibinfo {author} {\bibfnamefont {X.}~\bibnamefont {Ni}}, \bibinfo {author} {\bibfnamefont {E.}~\bibnamefont {Mazur}},\ and\ \bibinfo {author} {\bibfnamefont {E.~L.}\ \bibnamefont {Hu}},\ }\bibfield  {title} {\bibinfo {title} {Gan magic angle laser in a merged moiré photonic crystal},\ }\bibfield  {journal} {\bibinfo  {journal} {ACS Photonics}\ }\href {https://doi.org/10.1021/acsphotonics.3c01064} {10.1021/acsphotonics.3c01064} (\bibinfo {year} {2023})\BibitemShut {NoStop}%
\bibitem [{\citenamefont {Wang}\ \emph {et~al.}(2021)\citenamefont {Wang}, \citenamefont {Guo}, \citenamefont {Jin}, \citenamefont {Song},\ and\ \citenamefont {Fan}}]{RN67}%
  \BibitemOpen
  \bibfield  {author} {\bibinfo {author} {\bibfnamefont {H.}~\bibnamefont {Wang}}, \bibinfo {author} {\bibfnamefont {C.}~\bibnamefont {Guo}}, \bibinfo {author} {\bibfnamefont {W.}~\bibnamefont {Jin}}, \bibinfo {author} {\bibfnamefont {A.~Y.}\ \bibnamefont {Song}},\ and\ \bibinfo {author} {\bibfnamefont {S.}~\bibnamefont {Fan}},\ }\bibfield  {title} {\bibinfo {title} {Engineering arbitrarily oriented spatiotemporal optical vortices using transmission nodal lines},\ }\href@noop {} {\bibfield  {journal} {\bibinfo  {journal} {Optica}\ }\textbf {\bibinfo {volume} {8}},\ \bibinfo {pages} {966} (\bibinfo {year} {2021})}\BibitemShut {NoStop}%
\bibitem [{\citenamefont {Tang}\ \emph {et~al.}(2021{\natexlab{a}})\citenamefont {Tang}, \citenamefont {DeVault}, \citenamefont {Camayd-Muñoz}, \citenamefont {Liu}, \citenamefont {Jia}, \citenamefont {Du}, \citenamefont {Mello}, \citenamefont {Vulis}, \citenamefont {Li},\ and\ \citenamefont {Mazur}}]{RN26}%
  \BibitemOpen
  \bibfield  {author} {\bibinfo {author} {\bibfnamefont {H.}~\bibnamefont {Tang}}, \bibinfo {author} {\bibfnamefont {C.}~\bibnamefont {DeVault}}, \bibinfo {author} {\bibfnamefont {S.~A.}\ \bibnamefont {Camayd-Muñoz}}, \bibinfo {author} {\bibfnamefont {Y.}~\bibnamefont {Liu}}, \bibinfo {author} {\bibfnamefont {D.}~\bibnamefont {Jia}}, \bibinfo {author} {\bibfnamefont {F.}~\bibnamefont {Du}}, \bibinfo {author} {\bibfnamefont {O.}~\bibnamefont {Mello}}, \bibinfo {author} {\bibfnamefont {D.~I.}\ \bibnamefont {Vulis}}, \bibinfo {author} {\bibfnamefont {Y.}~\bibnamefont {Li}},\ and\ \bibinfo {author} {\bibfnamefont {E.}~\bibnamefont {Mazur}},\ }\bibfield  {title} {\bibinfo {title} {Low-loss zero-index materials},\ }\href@noop {} {\bibfield  {journal} {\bibinfo  {journal} {Nano Lett.}\ }\textbf {\bibinfo {volume} {21}},\ \bibinfo {pages} {914} (\bibinfo {year} {2021}{\natexlab{a}})}\BibitemShut {NoStop}%
\bibitem [{\citenamefont {Liu}\ \emph {et~al.}(2023)\citenamefont {Liu}, \citenamefont {Kan}, \citenamefont {Kumar}, \citenamefont {Komisar}, \citenamefont {Zhao},\ and\ \citenamefont {Bozhevolnyi}}]{RN10}%
  \BibitemOpen
  \bibfield  {author} {\bibinfo {author} {\bibfnamefont {X.}~\bibnamefont {Liu}}, \bibinfo {author} {\bibfnamefont {Y.}~\bibnamefont {Kan}}, \bibinfo {author} {\bibfnamefont {S.}~\bibnamefont {Kumar}}, \bibinfo {author} {\bibfnamefont {D.}~\bibnamefont {Komisar}}, \bibinfo {author} {\bibfnamefont {C.}~\bibnamefont {Zhao}},\ and\ \bibinfo {author} {\bibfnamefont {S.~I.}\ \bibnamefont {Bozhevolnyi}},\ }\bibfield  {title} {\bibinfo {title} {On-chip generation of single-photon circularly polarized single-mode vortex beams},\ }\href@noop {} {\bibfield  {journal} {\bibinfo  {journal} {Sci. Adv.}\ }\textbf {\bibinfo {volume} {9}},\ \bibinfo {pages} {eadh0725} (\bibinfo {year} {2023})}\BibitemShut {NoStop}%
\bibitem [{\citenamefont {Meng}\ \emph {et~al.}(2023)\citenamefont {Meng}, \citenamefont {Feng}, \citenamefont {Han}, \citenamefont {Xu}, \citenamefont {Mao}, \citenamefont {Zhang}, \citenamefont {Kim}, \citenamefont {Roh}, \citenamefont {Zhao}, \citenamefont {Kim}, \citenamefont {Yang}, \citenamefont {Lee}, \citenamefont {Yang}, \citenamefont {Qiu},\ and\ \citenamefont {Bae}}]{Rn75}%
  \BibitemOpen
  \bibfield  {author} {\bibinfo {author} {\bibfnamefont {Y.}~\bibnamefont {Meng}}, \bibinfo {author} {\bibfnamefont {J.}~\bibnamefont {Feng}}, \bibinfo {author} {\bibfnamefont {S.}~\bibnamefont {Han}}, \bibinfo {author} {\bibfnamefont {Z.}~\bibnamefont {Xu}}, \bibinfo {author} {\bibfnamefont {W.}~\bibnamefont {Mao}}, \bibinfo {author} {\bibfnamefont {T.}~\bibnamefont {Zhang}}, \bibinfo {author} {\bibfnamefont {J.~S.}\ \bibnamefont {Kim}}, \bibinfo {author} {\bibfnamefont {I.}~\bibnamefont {Roh}}, \bibinfo {author} {\bibfnamefont {Y.}~\bibnamefont {Zhao}}, \bibinfo {author} {\bibfnamefont {D.-H.}\ \bibnamefont {Kim}}, \bibinfo {author} {\bibfnamefont {Y.}~\bibnamefont {Yang}}, \bibinfo {author} {\bibfnamefont {J.-W.}\ \bibnamefont {Lee}}, \bibinfo {author} {\bibfnamefont {L.}~\bibnamefont {Yang}}, \bibinfo {author} {\bibfnamefont {C.-W.}\ \bibnamefont {Qiu}},\ and\ \bibinfo {author} {\bibfnamefont {S.-H.}\ \bibnamefont {Bae}},\ }\bibfield  {title} {\bibinfo {title} {Photonic van der waals integration from 2d
  materials to 3d nanomembranes},\ }\href {https://doi.org/10.1038/s41578-023-00558-w} {\bibfield  {journal} {\bibinfo  {journal} {Nature Reviews Materials}\ }\textbf {\bibinfo {volume} {8}},\ \bibinfo {pages} {498} (\bibinfo {year} {2023})}\BibitemShut {NoStop}%
\bibitem [{\citenamefont {Divitt}\ \emph {et~al.}(2019)\citenamefont {Divitt}, \citenamefont {Zhu}, \citenamefont {Zhang}, \citenamefont {Lezec},\ and\ \citenamefont {Agrawal}}]{RN7}%
  \BibitemOpen
  \bibfield  {author} {\bibinfo {author} {\bibfnamefont {S.}~\bibnamefont {Divitt}}, \bibinfo {author} {\bibfnamefont {W.}~\bibnamefont {Zhu}}, \bibinfo {author} {\bibfnamefont {C.}~\bibnamefont {Zhang}}, \bibinfo {author} {\bibfnamefont {H.~J.}\ \bibnamefont {Lezec}},\ and\ \bibinfo {author} {\bibfnamefont {A.}~\bibnamefont {Agrawal}},\ }\bibfield  {title} {\bibinfo {title} {Ultrafast optical pulse shaping using dielectric metasurfaces},\ }\href@noop {} {\bibfield  {journal} {\bibinfo  {journal} {Science}\ }\textbf {\bibinfo {volume} {364}},\ \bibinfo {pages} {890} (\bibinfo {year} {2019})}\BibitemShut {NoStop}%
\bibitem [{\citenamefont {Dorrah}\ \emph {et~al.}(2021)\citenamefont {Dorrah}, \citenamefont {Rubin}, \citenamefont {Tamagnone}, \citenamefont {Zaidi},\ and\ \citenamefont {Capasso}}]{RN46}%
  \BibitemOpen
  \bibfield  {author} {\bibinfo {author} {\bibfnamefont {A.~H.}\ \bibnamefont {Dorrah}}, \bibinfo {author} {\bibfnamefont {N.~A.}\ \bibnamefont {Rubin}}, \bibinfo {author} {\bibfnamefont {M.}~\bibnamefont {Tamagnone}}, \bibinfo {author} {\bibfnamefont {A.}~\bibnamefont {Zaidi}},\ and\ \bibinfo {author} {\bibfnamefont {F.}~\bibnamefont {Capasso}},\ }\bibfield  {title} {\bibinfo {title} {Structuring total angular momentum of light along the propagation direction with polarization-controlled meta-optics},\ }\href@noop {} {\bibfield  {journal} {\bibinfo  {journal} {Nat. Commun.}\ }\textbf {\bibinfo {volume} {12}},\ \bibinfo {pages} {6249} (\bibinfo {year} {2021})}\BibitemShut {NoStop}%
\bibitem [{\citenamefont {Ni}\ \emph {et~al.}(2023)\citenamefont {Ni}, \citenamefont {Ji}, \citenamefont {Wang}, \citenamefont {Liu}, \citenamefont {Hu}, \citenamefont {Chen}, \citenamefont {Li}, \citenamefont {Li}, \citenamefont {Chu}, \citenamefont {Wu},\ and\ \citenamefont {Qiu}}]{RN45}%
  \BibitemOpen
  \bibfield  {author} {\bibinfo {author} {\bibfnamefont {J.}~\bibnamefont {Ni}}, \bibinfo {author} {\bibfnamefont {S.}~\bibnamefont {Ji}}, \bibinfo {author} {\bibfnamefont {Z.}~\bibnamefont {Wang}}, \bibinfo {author} {\bibfnamefont {S.}~\bibnamefont {Liu}}, \bibinfo {author} {\bibfnamefont {Y.}~\bibnamefont {Hu}}, \bibinfo {author} {\bibfnamefont {Y.}~\bibnamefont {Chen}}, \bibinfo {author} {\bibfnamefont {J.}~\bibnamefont {Li}}, \bibinfo {author} {\bibfnamefont {X.}~\bibnamefont {Li}}, \bibinfo {author} {\bibfnamefont {J.}~\bibnamefont {Chu}}, \bibinfo {author} {\bibfnamefont {D.}~\bibnamefont {Wu}},\ and\ \bibinfo {author} {\bibfnamefont {C.-W.}\ \bibnamefont {Qiu}},\ }\bibfield  {title} {\bibinfo {title} {Unidirectional unpolarized luminescence emission via vortex excitation},\ }\href@noop {} {\bibfield  {journal} {\bibinfo  {journal} {Nat. Photonics}\ }\textbf {\bibinfo {volume} {17}},\ \bibinfo {pages} {601} (\bibinfo {year} {2023})}\BibitemShut {NoStop}%
\bibitem [{\citenamefont {Ren}\ \emph {et~al.}(2019)\citenamefont {Ren}, \citenamefont {Briere}, \citenamefont {Fang}, \citenamefont {Ni}, \citenamefont {Sawant}, \citenamefont {Héron}, \citenamefont {Chenot}, \citenamefont {Vézian}, \citenamefont {Damilano}, \citenamefont {Brändli}, \citenamefont {Maier},\ and\ \citenamefont {Genevet}}]{RN39}%
  \BibitemOpen
  \bibfield  {author} {\bibinfo {author} {\bibfnamefont {H.}~\bibnamefont {Ren}}, \bibinfo {author} {\bibfnamefont {G.}~\bibnamefont {Briere}}, \bibinfo {author} {\bibfnamefont {X.}~\bibnamefont {Fang}}, \bibinfo {author} {\bibfnamefont {P.}~\bibnamefont {Ni}}, \bibinfo {author} {\bibfnamefont {R.}~\bibnamefont {Sawant}}, \bibinfo {author} {\bibfnamefont {S.}~\bibnamefont {Héron}}, \bibinfo {author} {\bibfnamefont {S.}~\bibnamefont {Chenot}}, \bibinfo {author} {\bibfnamefont {S.}~\bibnamefont {Vézian}}, \bibinfo {author} {\bibfnamefont {B.}~\bibnamefont {Damilano}}, \bibinfo {author} {\bibfnamefont {V.}~\bibnamefont {Brändli}}, \bibinfo {author} {\bibfnamefont {S.~A.}\ \bibnamefont {Maier}},\ and\ \bibinfo {author} {\bibfnamefont {P.}~\bibnamefont {Genevet}},\ }\bibfield  {title} {\bibinfo {title} {Metasurface orbital angular momentum holography},\ }\href@noop {} {\bibfield  {journal} {\bibinfo  {journal} {Nat. Commun.}\ }\textbf {\bibinfo {volume} {10}},\ \bibinfo {pages} {2986} (\bibinfo {year}
  {2019})}\BibitemShut {NoStop}%
\bibitem [{\citenamefont {Zeng}\ \emph {et~al.}(2021)\citenamefont {Zeng}, \citenamefont {Hu}, \citenamefont {Liu}, \citenamefont {Tang},\ and\ \citenamefont {Qiu}}]{RN57}%
  \BibitemOpen
  \bibfield  {author} {\bibinfo {author} {\bibfnamefont {Y.}~\bibnamefont {Zeng}}, \bibinfo {author} {\bibfnamefont {G.}~\bibnamefont {Hu}}, \bibinfo {author} {\bibfnamefont {K.}~\bibnamefont {Liu}}, \bibinfo {author} {\bibfnamefont {Z.}~\bibnamefont {Tang}},\ and\ \bibinfo {author} {\bibfnamefont {C.-W.}\ \bibnamefont {Qiu}},\ }\bibfield  {title} {\bibinfo {title} {Dynamics of topological polarization singularity in momentum space},\ }\href@noop {} {\bibfield  {journal} {\bibinfo  {journal} {Phys. Rev. Lett.}\ }\textbf {\bibinfo {volume} {127}} (\bibinfo {year} {2021})}\BibitemShut {NoStop}%
\bibitem [{\citenamefont {Zhang}\ \emph {et~al.}(2022)\citenamefont {Zhang}, \citenamefont {Liu}, \citenamefont {Han}, \citenamefont {Kivshar},\ and\ \citenamefont {Song}}]{RN44}%
  \BibitemOpen
  \bibfield  {author} {\bibinfo {author} {\bibfnamefont {X.}~\bibnamefont {Zhang}}, \bibinfo {author} {\bibfnamefont {Y.}~\bibnamefont {Liu}}, \bibinfo {author} {\bibfnamefont {J.}~\bibnamefont {Han}}, \bibinfo {author} {\bibfnamefont {Y.}~\bibnamefont {Kivshar}},\ and\ \bibinfo {author} {\bibfnamefont {Q.}~\bibnamefont {Song}},\ }\bibfield  {title} {\bibinfo {title} {Chiral emission from resonant metasurfaces},\ }\href@noop {} {\bibfield  {journal} {\bibinfo  {journal} {Science}\ }\textbf {\bibinfo {volume} {377}},\ \bibinfo {pages} {1215} (\bibinfo {year} {2022})}\BibitemShut {NoStop}%
\bibitem [{\citenamefont {Chen}\ \emph {et~al.}(2021{\natexlab{b}})\citenamefont {Chen}, \citenamefont {Lin}, \citenamefont {Chen}, \citenamefont {Low}, \citenamefont {Chen},\ and\ \citenamefont {Dai}}]{RN24}%
  \BibitemOpen
  \bibfield  {author} {\bibinfo {author} {\bibfnamefont {J.}~\bibnamefont {Chen}}, \bibinfo {author} {\bibfnamefont {X.}~\bibnamefont {Lin}}, \bibinfo {author} {\bibfnamefont {M.}~\bibnamefont {Chen}}, \bibinfo {author} {\bibfnamefont {T.}~\bibnamefont {Low}}, \bibinfo {author} {\bibfnamefont {H.}~\bibnamefont {Chen}},\ and\ \bibinfo {author} {\bibfnamefont {S.}~\bibnamefont {Dai}},\ }\bibfield  {title} {\bibinfo {title} {A perspective of twisted photonic structures},\ }\href@noop {} {\bibfield  {journal} {\bibinfo  {journal} {Appl. Phys. Lett.}\ }\textbf {\bibinfo {volume} {119}},\ \bibinfo {pages} {240501} (\bibinfo {year} {2021}{\natexlab{b}})}\BibitemShut {NoStop}%
\bibitem [{\citenamefont {Guan}\ \emph {et~al.}(2023)\citenamefont {Guan}, \citenamefont {Hu}, \citenamefont {Wang}, \citenamefont {Tan}, \citenamefont {Schatz},\ and\ \citenamefont {Odom}}]{RN6}%
  \BibitemOpen
  \bibfield  {author} {\bibinfo {author} {\bibfnamefont {J.}~\bibnamefont {Guan}}, \bibinfo {author} {\bibfnamefont {J.}~\bibnamefont {Hu}}, \bibinfo {author} {\bibfnamefont {Y.}~\bibnamefont {Wang}}, \bibinfo {author} {\bibfnamefont {M.~J.~H.}\ \bibnamefont {Tan}}, \bibinfo {author} {\bibfnamefont {G.~C.}\ \bibnamefont {Schatz}},\ and\ \bibinfo {author} {\bibfnamefont {T.~W.}\ \bibnamefont {Odom}},\ }\bibfield  {title} {\bibinfo {title} {Far-field coupling between moiré photonic lattices},\ }\href@noop {} {\bibfield  {journal} {\bibinfo  {journal} {Nat. Nanotechnol.}\ } (\bibinfo {year} {2023})}\BibitemShut {NoStop}%
\bibitem [{\citenamefont {Lou}\ and\ \citenamefont {Fan}(2022)}]{RN20}%
  \BibitemOpen
  \bibfield  {author} {\bibinfo {author} {\bibfnamefont {B.}~\bibnamefont {Lou}}\ and\ \bibinfo {author} {\bibfnamefont {S.}~\bibnamefont {Fan}},\ }\bibfield  {title} {\bibinfo {title} {Tunable frequency filter based on twisted bilayer photonic crystal slabs},\ }\href@noop {} {\bibfield  {journal} {\bibinfo  {journal} {ACS Photonics}\ }\textbf {\bibinfo {volume} {9}},\ \bibinfo {pages} {800} (\bibinfo {year} {2022})}\BibitemShut {NoStop}%
\bibitem [{\citenamefont {Lou}\ \emph {et~al.}(2022)\citenamefont {Lou}, \citenamefont {Wang}, \citenamefont {Rodríguez}, \citenamefont {Cappelli},\ and\ \citenamefont {Fan}}]{RN16}%
  \BibitemOpen
  \bibfield  {author} {\bibinfo {author} {\bibfnamefont {B.}~\bibnamefont {Lou}}, \bibinfo {author} {\bibfnamefont {B.}~\bibnamefont {Wang}}, \bibinfo {author} {\bibfnamefont {J.~A.}\ \bibnamefont {Rodríguez}}, \bibinfo {author} {\bibfnamefont {M.}~\bibnamefont {Cappelli}},\ and\ \bibinfo {author} {\bibfnamefont {S.}~\bibnamefont {Fan}},\ }\bibfield  {title} {\bibinfo {title} {Tunable guided resonance in twisted bilayer photonic crystal},\ }\href@noop {} {\bibfield  {journal} {\bibinfo  {journal} {Sci Adv}\ }\textbf {\bibinfo {volume} {8}},\ \bibinfo {pages} {eadd4339} (\bibinfo {year} {2022})}\BibitemShut {NoStop}%
\bibitem [{\citenamefont {Lou}\ \emph {et~al.}(2021)\citenamefont {Lou}, \citenamefont {Zhao}, \citenamefont {Minkov}, \citenamefont {Guo}, \citenamefont {Orenstein},\ and\ \citenamefont {Fan}}]{RN25}%
  \BibitemOpen
  \bibfield  {author} {\bibinfo {author} {\bibfnamefont {B.}~\bibnamefont {Lou}}, \bibinfo {author} {\bibfnamefont {N.}~\bibnamefont {Zhao}}, \bibinfo {author} {\bibfnamefont {M.}~\bibnamefont {Minkov}}, \bibinfo {author} {\bibfnamefont {C.}~\bibnamefont {Guo}}, \bibinfo {author} {\bibfnamefont {M.}~\bibnamefont {Orenstein}},\ and\ \bibinfo {author} {\bibfnamefont {S.}~\bibnamefont {Fan}},\ }\bibfield  {title} {\bibinfo {title} {Theory for twisted bilayer photonic crystal slabs},\ }\href@noop {} {\bibfield  {journal} {\bibinfo  {journal} {Phys. Rev. Lett.}\ }\textbf {\bibinfo {volume} {126}},\ \bibinfo {pages} {136101} (\bibinfo {year} {2021})}\BibitemShut {NoStop}%
\bibitem [{\citenamefont {Lu}\ and\ \citenamefont {Lee}(2009)}]{RN28}%
  \BibitemOpen
  \bibfield  {author} {\bibinfo {author} {\bibfnamefont {T.-W.}\ \bibnamefont {Lu}}\ and\ \bibinfo {author} {\bibfnamefont {P.-T.}\ \bibnamefont {Lee}},\ }\bibfield  {title} {\bibinfo {title} {Ultra-high sensitivity optical stress sensor based on double-layered photonic crystal microcavity},\ }\href@noop {} {\bibfield  {journal} {\bibinfo  {journal} {Opt. Express}\ }\textbf {\bibinfo {volume} {17}},\ \bibinfo {pages} {1518} (\bibinfo {year} {2009})}\BibitemShut {NoStop}%
\bibitem [{\citenamefont {Nguyen}\ \emph {et~al.}(2023)\citenamefont {Nguyen}, \citenamefont {Devescovi}, \citenamefont {Nguyen}, \citenamefont {Nguyen},\ and\ \citenamefont {Bercioux}}]{PhysRevLett.131.053602}%
  \BibitemOpen
  \bibfield  {author} {\bibinfo {author} {\bibfnamefont {D.-H.-M.}\ \bibnamefont {Nguyen}}, \bibinfo {author} {\bibfnamefont {C.}~\bibnamefont {Devescovi}}, \bibinfo {author} {\bibfnamefont {D.~X.}\ \bibnamefont {Nguyen}}, \bibinfo {author} {\bibfnamefont {H.~S.}\ \bibnamefont {Nguyen}},\ and\ \bibinfo {author} {\bibfnamefont {D.}~\bibnamefont {Bercioux}},\ }\bibfield  {title} {\bibinfo {title} {Fermi arc reconstruction in synthetic photonic lattice},\ }\href {https://doi.org/10.1103/PhysRevLett.131.053602} {\bibfield  {journal} {\bibinfo  {journal} {Phys. Rev. Lett.}\ }\textbf {\bibinfo {volume} {131}},\ \bibinfo {pages} {053602} (\bibinfo {year} {2023})}\BibitemShut {NoStop}%
\bibitem [{\citenamefont {Nguyen}\ \emph {et~al.}(2022)\citenamefont {Nguyen}, \citenamefont {Letartre}, \citenamefont {Drouard}, \citenamefont {Viktorovitch}, \citenamefont {Nguyen},\ and\ \citenamefont {Nguyen}}]{RN22}%
  \BibitemOpen
  \bibfield  {author} {\bibinfo {author} {\bibfnamefont {D.~X.}\ \bibnamefont {Nguyen}}, \bibinfo {author} {\bibfnamefont {X.}~\bibnamefont {Letartre}}, \bibinfo {author} {\bibfnamefont {E.}~\bibnamefont {Drouard}}, \bibinfo {author} {\bibfnamefont {P.}~\bibnamefont {Viktorovitch}}, \bibinfo {author} {\bibfnamefont {H.~C.}\ \bibnamefont {Nguyen}},\ and\ \bibinfo {author} {\bibfnamefont {H.~S.}\ \bibnamefont {Nguyen}},\ }\bibfield  {title} {\bibinfo {title} {Magic configurations in moiré superlattice of bilayer photonic crystals: Almost-perfect flatbands and unconventional localization},\ }\href@noop {} {\bibfield  {journal} {\bibinfo  {journal} {Phys. Rev. Research}\ }\textbf {\bibinfo {volume} {4}} (\bibinfo {year} {2022})}\BibitemShut {NoStop}%
\bibitem [{\citenamefont {Qin}\ \emph {et~al.}(2023)\citenamefont {Qin}, \citenamefont {Su}, \citenamefont {Liu}, \citenamefont {Zeng}, \citenamefont {Tang}, \citenamefont {Li}, \citenamefont {Shi}, \citenamefont {Huang}, \citenamefont {Qiu},\ and\ \citenamefont {Song}}]{RN62}%
  \BibitemOpen
  \bibfield  {author} {\bibinfo {author} {\bibfnamefont {H.}~\bibnamefont {Qin}}, \bibinfo {author} {\bibfnamefont {Z.}~\bibnamefont {Su}}, \bibinfo {author} {\bibfnamefont {M.}~\bibnamefont {Liu}}, \bibinfo {author} {\bibfnamefont {Y.}~\bibnamefont {Zeng}}, \bibinfo {author} {\bibfnamefont {M.-C.}\ \bibnamefont {Tang}}, \bibinfo {author} {\bibfnamefont {M.}~\bibnamefont {Li}}, \bibinfo {author} {\bibfnamefont {Y.}~\bibnamefont {Shi}}, \bibinfo {author} {\bibfnamefont {W.}~\bibnamefont {Huang}}, \bibinfo {author} {\bibfnamefont {C.-W.}\ \bibnamefont {Qiu}},\ and\ \bibinfo {author} {\bibfnamefont {Q.}~\bibnamefont {Song}},\ }\bibfield  {title} {\bibinfo {title} {Arbitrarily polarized bound states in the continuum with twisted photonic crystal slabs},\ }\href@noop {} {\bibfield  {journal} {\bibinfo  {journal} {Light Sci. Appl.}\ }\textbf {\bibinfo {volume} {12}},\ \bibinfo {pages} {66} (\bibinfo {year} {2023})}\BibitemShut {NoStop}%
\bibitem [{\citenamefont {Shuai}\ \emph {et~al.}(2013{\natexlab{a}})\citenamefont {Shuai}, \citenamefont {Zhao}, \citenamefont {Singh~Chadha}, \citenamefont {Seo}, \citenamefont {Yang}, \citenamefont {Fan}, \citenamefont {Ma},\ and\ \citenamefont {Zhou}}]{RN33}%
  \BibitemOpen
  \bibfield  {author} {\bibinfo {author} {\bibfnamefont {Y.}~\bibnamefont {Shuai}}, \bibinfo {author} {\bibfnamefont {D.}~\bibnamefont {Zhao}}, \bibinfo {author} {\bibfnamefont {A.}~\bibnamefont {Singh~Chadha}}, \bibinfo {author} {\bibfnamefont {J.-H.}\ \bibnamefont {Seo}}, \bibinfo {author} {\bibfnamefont {H.}~\bibnamefont {Yang}}, \bibinfo {author} {\bibfnamefont {S.}~\bibnamefont {Fan}}, \bibinfo {author} {\bibfnamefont {Z.}~\bibnamefont {Ma}},\ and\ \bibinfo {author} {\bibfnamefont {W.}~\bibnamefont {Zhou}},\ }\bibfield  {title} {\bibinfo {title} {Coupled double-layer fano resonance photonic crystal filters with lattice-displacement},\ }\href@noop {} {\bibfield  {journal} {\bibinfo  {journal} {Appl. Phys. Lett.}\ }\textbf {\bibinfo {volume} {103}},\ \bibinfo {pages} {241106} (\bibinfo {year} {2013}{\natexlab{a}})}\BibitemShut {NoStop}%
\bibitem [{\citenamefont {Shuai}\ \emph {et~al.}(2013{\natexlab{b}})\citenamefont {Shuai}, \citenamefont {Zhao}, \citenamefont {Tian}, \citenamefont {Seo}, \citenamefont {Plant}, \citenamefont {Ma}, \citenamefont {Fan},\ and\ \citenamefont {Zhou}}]{RN36}%
  \BibitemOpen
  \bibfield  {author} {\bibinfo {author} {\bibfnamefont {Y.}~\bibnamefont {Shuai}}, \bibinfo {author} {\bibfnamefont {D.}~\bibnamefont {Zhao}}, \bibinfo {author} {\bibfnamefont {Z.}~\bibnamefont {Tian}}, \bibinfo {author} {\bibfnamefont {J.-H.}\ \bibnamefont {Seo}}, \bibinfo {author} {\bibfnamefont {D.~V.}\ \bibnamefont {Plant}}, \bibinfo {author} {\bibfnamefont {Z.}~\bibnamefont {Ma}}, \bibinfo {author} {\bibfnamefont {S.}~\bibnamefont {Fan}},\ and\ \bibinfo {author} {\bibfnamefont {W.}~\bibnamefont {Zhou}},\ }\bibfield  {title} {\bibinfo {title} {Double-layer fano resonance photonic crystal filters},\ }\href@noop {} {\bibfield  {journal} {\bibinfo  {journal} {Opt. Express}\ }\textbf {\bibinfo {volume} {21}},\ \bibinfo {pages} {24582} (\bibinfo {year} {2013}{\natexlab{b}})}\BibitemShut {NoStop}%
\bibitem [{\citenamefont {Tang}\ \emph {et~al.}(2021{\natexlab{b}})\citenamefont {Tang}, \citenamefont {Du}, \citenamefont {Carr}, \citenamefont {DeVault}, \citenamefont {Mello},\ and\ \citenamefont {Mazur}}]{RN23}%
  \BibitemOpen
  \bibfield  {author} {\bibinfo {author} {\bibfnamefont {H.}~\bibnamefont {Tang}}, \bibinfo {author} {\bibfnamefont {F.}~\bibnamefont {Du}}, \bibinfo {author} {\bibfnamefont {S.}~\bibnamefont {Carr}}, \bibinfo {author} {\bibfnamefont {C.}~\bibnamefont {DeVault}}, \bibinfo {author} {\bibfnamefont {O.}~\bibnamefont {Mello}},\ and\ \bibinfo {author} {\bibfnamefont {E.}~\bibnamefont {Mazur}},\ }\bibfield  {title} {\bibinfo {title} {Modeling the optical properties of twisted bilayer photonic crystals},\ }\href@noop {} {\bibfield  {journal} {\bibinfo  {journal} {Light Sci Appl}\ }\textbf {\bibinfo {volume} {10}},\ \bibinfo {pages} {157} (\bibinfo {year} {2021}{\natexlab{b}})}\BibitemShut {NoStop}%
\bibitem [{\citenamefont {Tang}\ \emph {et~al.}(2022)\citenamefont {Tang}, \citenamefont {Ni}, \citenamefont {Du}, \citenamefont {Srikrishna},\ and\ \citenamefont {Mazur}}]{RN17}%
  \BibitemOpen
  \bibfield  {author} {\bibinfo {author} {\bibfnamefont {H.}~\bibnamefont {Tang}}, \bibinfo {author} {\bibfnamefont {X.}~\bibnamefont {Ni}}, \bibinfo {author} {\bibfnamefont {F.}~\bibnamefont {Du}}, \bibinfo {author} {\bibfnamefont {V.}~\bibnamefont {Srikrishna}},\ and\ \bibinfo {author} {\bibfnamefont {E.}~\bibnamefont {Mazur}},\ }\bibfield  {title} {\bibinfo {title} {On-chip light trapping in bilayer moiré photonic crystal slabs},\ }\href@noop {} {\bibfield  {journal} {\bibinfo  {journal} {Appl. Phys. Lett.}\ }\textbf {\bibinfo {volume} {121}},\ \bibinfo {pages} {231702} (\bibinfo {year} {2022})}\BibitemShut {NoStop}%
\bibitem [{\citenamefont {Yu}\ \emph {et~al.}(2023)\citenamefont {Yu}, \citenamefont {Li}, \citenamefont {Wang}, \citenamefont {Leykam}, \citenamefont {Yuan},\ and\ \citenamefont {Chen}}]{RN9}%
  \BibitemOpen
  \bibfield  {author} {\bibinfo {author} {\bibfnamefont {D.}~\bibnamefont {Yu}}, \bibinfo {author} {\bibfnamefont {G.}~\bibnamefont {Li}}, \bibinfo {author} {\bibfnamefont {L.}~\bibnamefont {Wang}}, \bibinfo {author} {\bibfnamefont {D.}~\bibnamefont {Leykam}}, \bibinfo {author} {\bibfnamefont {L.}~\bibnamefont {Yuan}},\ and\ \bibinfo {author} {\bibfnamefont {X.}~\bibnamefont {Chen}},\ }\bibfield  {title} {\bibinfo {title} {Moiré lattice in one-dimensional synthetic frequency dimension},\ }\href@noop {} {\bibfield  {journal} {\bibinfo  {journal} {Phys. Rev. Lett.}\ }\textbf {\bibinfo {volume} {130}},\ \bibinfo {pages} {143801} (\bibinfo {year} {2023})}\BibitemShut {NoStop}%
\bibitem [{\citenamefont {Salakhova}\ \emph {et~al.}(2023)\citenamefont {Salakhova}, \citenamefont {Fradkin}, \citenamefont {Dyakov},\ and\ \citenamefont {Gippius}}]{PhysRevB.107.155402}%
  \BibitemOpen
  \bibfield  {author} {\bibinfo {author} {\bibfnamefont {N.~S.}\ \bibnamefont {Salakhova}}, \bibinfo {author} {\bibfnamefont {I.~M.}\ \bibnamefont {Fradkin}}, \bibinfo {author} {\bibfnamefont {S.~A.}\ \bibnamefont {Dyakov}},\ and\ \bibinfo {author} {\bibfnamefont {N.~A.}\ \bibnamefont {Gippius}},\ }\bibfield  {title} {\bibinfo {title} {Twist-tunable moir\'e optical resonances},\ }\href {https://doi.org/10.1103/PhysRevB.107.155402} {\bibfield  {journal} {\bibinfo  {journal} {Phys. Rev. B}\ }\textbf {\bibinfo {volume} {107}},\ \bibinfo {pages} {155402} (\bibinfo {year} {2023})}\BibitemShut {NoStop}%
\bibitem [{\citenamefont {Saadi}\ \emph {et~al.}(2023)\citenamefont {Saadi}, \citenamefont {Nguyen}, \citenamefont {Cueff}, \citenamefont {Ferrier}, \citenamefont {Letartre},\ and\ \citenamefont {Callard}}]{saadi2023supercells}%
  \BibitemOpen
  \bibfield  {author} {\bibinfo {author} {\bibfnamefont {C.}~\bibnamefont {Saadi}}, \bibinfo {author} {\bibfnamefont {H.~S.}\ \bibnamefont {Nguyen}}, \bibinfo {author} {\bibfnamefont {S.}~\bibnamefont {Cueff}}, \bibinfo {author} {\bibfnamefont {L.}~\bibnamefont {Ferrier}}, \bibinfo {author} {\bibfnamefont {X.}~\bibnamefont {Letartre}},\ and\ \bibinfo {author} {\bibfnamefont {S.}~\bibnamefont {Callard}},\ }\href@noop {} {\bibinfo {title} {How many supercells are required to achieve unconventional light confinement effects in moir\'e photonic lattices?}} (\bibinfo {year} {2023}),\ \Eprint {https://arxiv.org/abs/2307.05525} {arXiv:2307.05525 [physics.optics]} \BibitemShut {NoStop}%
\bibitem [{\citenamefont {Arbabi}\ \emph {et~al.}(2018)\citenamefont {Arbabi}, \citenamefont {Arbabi}, \citenamefont {Kamali}, \citenamefont {Horie}, \citenamefont {Faraji-Dana},\ and\ \citenamefont {Faraon}}]{RN31}%
  \BibitemOpen
  \bibfield  {author} {\bibinfo {author} {\bibfnamefont {E.}~\bibnamefont {Arbabi}}, \bibinfo {author} {\bibfnamefont {A.}~\bibnamefont {Arbabi}}, \bibinfo {author} {\bibfnamefont {S.~M.}\ \bibnamefont {Kamali}}, \bibinfo {author} {\bibfnamefont {Y.}~\bibnamefont {Horie}}, \bibinfo {author} {\bibfnamefont {M.}~\bibnamefont {Faraji-Dana}},\ and\ \bibinfo {author} {\bibfnamefont {A.}~\bibnamefont {Faraon}},\ }\bibfield  {title} {\bibinfo {title} {Mems-tunable dielectric metasurface lens},\ }\href@noop {} {\bibfield  {journal} {\bibinfo  {journal} {Nat. Commun.}\ }\textbf {\bibinfo {volume} {9}},\ \bibinfo {pages} {812} (\bibinfo {year} {2018})}\BibitemShut {NoStop}%
\bibitem [{\citenamefont {Petruzzella}\ \emph {et~al.}(2016)\citenamefont {Petruzzella}, \citenamefont {La~China}, \citenamefont {Intonti}, \citenamefont {Caselli}, \citenamefont {De~Pas}, \citenamefont {van Otten}, \citenamefont {Gurioli},\ and\ \citenamefont {Fiore}}]{RN29}%
  \BibitemOpen
  \bibfield  {author} {\bibinfo {author} {\bibfnamefont {M.}~\bibnamefont {Petruzzella}}, \bibinfo {author} {\bibfnamefont {F.}~\bibnamefont {La~China}}, \bibinfo {author} {\bibfnamefont {F.}~\bibnamefont {Intonti}}, \bibinfo {author} {\bibfnamefont {N.}~\bibnamefont {Caselli}}, \bibinfo {author} {\bibfnamefont {M.}~\bibnamefont {De~Pas}}, \bibinfo {author} {\bibfnamefont {F.~W.~M.}\ \bibnamefont {van Otten}}, \bibinfo {author} {\bibfnamefont {M.}~\bibnamefont {Gurioli}},\ and\ \bibinfo {author} {\bibfnamefont {A.}~\bibnamefont {Fiore}},\ }\bibfield  {title} {\bibinfo {title} {Nanoscale mechanical actuation and near-field read-out of photonic crystal molecules},\ }\href@noop {} {\bibfield  {journal} {\bibinfo  {journal} {Phys. Rev. B Condens. Matter}\ }\textbf {\bibinfo {volume} {94}} (\bibinfo {year} {2016})}\BibitemShut {NoStop}%
\bibitem [{\citenamefont {Roy}\ \emph {et~al.}(2018)\citenamefont {Roy}, \citenamefont {Zhang}, \citenamefont {Jung}, \citenamefont {Troccoli}, \citenamefont {Capasso},\ and\ \citenamefont {Lopez}}]{RN43}%
  \BibitemOpen
  \bibfield  {author} {\bibinfo {author} {\bibfnamefont {T.}~\bibnamefont {Roy}}, \bibinfo {author} {\bibfnamefont {S.}~\bibnamefont {Zhang}}, \bibinfo {author} {\bibfnamefont {I.~W.}\ \bibnamefont {Jung}}, \bibinfo {author} {\bibfnamefont {M.}~\bibnamefont {Troccoli}}, \bibinfo {author} {\bibfnamefont {F.}~\bibnamefont {Capasso}},\ and\ \bibinfo {author} {\bibfnamefont {D.}~\bibnamefont {Lopez}},\ }\bibfield  {title} {\bibinfo {title} {Dynamic metasurface lens based on mems technology},\ }\href@noop {} {\bibfield  {journal} {\bibinfo  {journal} {APL Photonics}\ }\textbf {\bibinfo {volume} {3}},\ \bibinfo {pages} {021302} (\bibinfo {year} {2018})}\BibitemShut {NoStop}%
\bibitem [{\citenamefont {Joannopoulos}\ \emph {et~al.}(2008)\citenamefont {Joannopoulos}, \citenamefont {Johnson}, \citenamefont {Winn},\ and\ \citenamefont {Meade}}]{RN21}%
  \BibitemOpen
  \bibfield  {author} {\bibinfo {author} {\bibfnamefont {J.~D.}\ \bibnamefont {Joannopoulos}}, \bibinfo {author} {\bibfnamefont {S.~G.}\ \bibnamefont {Johnson}}, \bibinfo {author} {\bibfnamefont {J.~N.}\ \bibnamefont {Winn}},\ and\ \bibinfo {author} {\bibfnamefont {R.~D.}\ \bibnamefont {Meade}},\ }\href@noop {} {\emph {\bibinfo {title} {Photonic Crystals: Molding the Flow of Light (Second Edition)}}}\ (\bibinfo  {publisher} {Princeton University Press},\ \bibinfo {year} {2008})\BibitemShut {NoStop}%
\bibitem [{\citenamefont {Fan}\ and\ \citenamefont {Joannopoulos}(2002)}]{PhysRevB.65.235112}%
  \BibitemOpen
  \bibfield  {author} {\bibinfo {author} {\bibfnamefont {S.}~\bibnamefont {Fan}}\ and\ \bibinfo {author} {\bibfnamefont {J.~D.}\ \bibnamefont {Joannopoulos}},\ }\bibfield  {title} {\bibinfo {title} {Analysis of guided resonances in photonic crystal slabs},\ }\href {https://doi.org/10.1103/PhysRevB.65.235112} {\bibfield  {journal} {\bibinfo  {journal} {Phys. Rev. B}\ }\textbf {\bibinfo {volume} {65}},\ \bibinfo {pages} {235112} (\bibinfo {year} {2002})}\BibitemShut {NoStop}%
\bibitem [{\citenamefont {Tang}\ \emph {et~al.}(2023)\citenamefont {Tang}, \citenamefont {Lou}, \citenamefont {Du}, \citenamefont {Zhang}, \citenamefont {Ni}, \citenamefont {Xu}, \citenamefont {Jin}, \citenamefont {Fan},\ and\ \citenamefont {Mazur}}]{RN60}%
  \BibitemOpen
  \bibfield  {author} {\bibinfo {author} {\bibfnamefont {H.}~\bibnamefont {Tang}}, \bibinfo {author} {\bibfnamefont {B.}~\bibnamefont {Lou}}, \bibinfo {author} {\bibfnamefont {F.}~\bibnamefont {Du}}, \bibinfo {author} {\bibfnamefont {M.}~\bibnamefont {Zhang}}, \bibinfo {author} {\bibfnamefont {X.}~\bibnamefont {Ni}}, \bibinfo {author} {\bibfnamefont {W.}~\bibnamefont {Xu}}, \bibinfo {author} {\bibfnamefont {R.}~\bibnamefont {Jin}}, \bibinfo {author} {\bibfnamefont {S.}~\bibnamefont {Fan}},\ and\ \bibinfo {author} {\bibfnamefont {E.}~\bibnamefont {Mazur}},\ }\bibfield  {title} {\bibinfo {title} {Experimental probe of twist angle-dependent band structure of on-chip optical bilayer photonic crystal},\ }\href@noop {} {\bibfield  {journal} {\bibinfo  {journal} {Sci. Adv.}\ }\textbf {\bibinfo {volume} {9}},\ \bibinfo {pages} {eadh8498} (\bibinfo {year} {2023})}\BibitemShut {NoStop}%
\bibitem [{\citenamefont {Yves}\ \emph {et~al.}(2022)\citenamefont {Yves}, \citenamefont {Peng},\ and\ \citenamefont {Alù}}]{RN5}%
  \BibitemOpen
  \bibfield  {author} {\bibinfo {author} {\bibfnamefont {S.}~\bibnamefont {Yves}}, \bibinfo {author} {\bibfnamefont {Y.-G.}\ \bibnamefont {Peng}},\ and\ \bibinfo {author} {\bibfnamefont {A.}~\bibnamefont {Alù}},\ }\bibfield  {title} {\bibinfo {title} {Topological lifshitz transition in twisted hyperbolic acoustic metasurfaces},\ }\href@noop {} {\bibfield  {journal} {\bibinfo  {journal} {Appl. Phys. Lett.}\ }\textbf {\bibinfo {volume} {121}},\ \bibinfo {pages} {122201} (\bibinfo {year} {2022})}\BibitemShut {NoStop}%
\bibitem [{\citenamefont {Suh}\ \emph {et~al.}(2003)\citenamefont {Suh}, \citenamefont {Yanik}, \citenamefont {Solgaard},\ and\ \citenamefont {Fan}}]{10.1063/1.1563739}%
  \BibitemOpen
  \bibfield  {author} {\bibinfo {author} {\bibfnamefont {W.}~\bibnamefont {Suh}}, \bibinfo {author} {\bibfnamefont {M.~F.}\ \bibnamefont {Yanik}}, \bibinfo {author} {\bibfnamefont {O.}~\bibnamefont {Solgaard}},\ and\ \bibinfo {author} {\bibfnamefont {S.}~\bibnamefont {Fan}},\ }\bibfield  {title} {\bibinfo {title} {{Displacement-sensitive photonic crystal structures based on guided resonance in photonic crystal slabs}},\ }\href {https://doi.org/10.1063/1.1563739} {\bibfield  {journal} {\bibinfo  {journal} {Applied Physics Letters}\ }\textbf {\bibinfo {volume} {82}},\ \bibinfo {pages} {1999} (\bibinfo {year} {2003})}\BibitemShut {NoStop}%
\bibitem [{\citenamefont {Zhen}\ \emph {et~al.}(2014)\citenamefont {Zhen}, \citenamefont {Hsu}, \citenamefont {Lu}, \citenamefont {Stone},\ and\ \citenamefont {Soljačić}}]{RN13}%
  \BibitemOpen
  \bibfield  {author} {\bibinfo {author} {\bibfnamefont {B.}~\bibnamefont {Zhen}}, \bibinfo {author} {\bibfnamefont {C.~W.}\ \bibnamefont {Hsu}}, \bibinfo {author} {\bibfnamefont {L.}~\bibnamefont {Lu}}, \bibinfo {author} {\bibfnamefont {A.~D.}\ \bibnamefont {Stone}},\ and\ \bibinfo {author} {\bibfnamefont {M.}~\bibnamefont {Soljačić}},\ }\bibfield  {title} {\bibinfo {title} {Topological nature of optical bound states in the continuum},\ }\href@noop {} {\bibfield  {journal} {\bibinfo  {journal} {Phys. Rev. Lett.}\ }\textbf {\bibinfo {volume} {113}},\ \bibinfo {pages} {257401} (\bibinfo {year} {2014})}\BibitemShut {NoStop}%
\bibitem [{\citenamefont {Wang}\ \emph {et~al.}(2020)\citenamefont {Wang}, \citenamefont {Liu}, \citenamefont {Zhao}, \citenamefont {Wang}, \citenamefont {Zhang}, \citenamefont {Chen}, \citenamefont {Guan}, \citenamefont {Liu}, \citenamefont {Shi},\ and\ \citenamefont {Zi}}]{RN59}%
  \BibitemOpen
  \bibfield  {author} {\bibinfo {author} {\bibfnamefont {B.}~\bibnamefont {Wang}}, \bibinfo {author} {\bibfnamefont {W.}~\bibnamefont {Liu}}, \bibinfo {author} {\bibfnamefont {M.}~\bibnamefont {Zhao}}, \bibinfo {author} {\bibfnamefont {J.}~\bibnamefont {Wang}}, \bibinfo {author} {\bibfnamefont {Y.}~\bibnamefont {Zhang}}, \bibinfo {author} {\bibfnamefont {A.}~\bibnamefont {Chen}}, \bibinfo {author} {\bibfnamefont {F.}~\bibnamefont {Guan}}, \bibinfo {author} {\bibfnamefont {X.}~\bibnamefont {Liu}}, \bibinfo {author} {\bibfnamefont {L.}~\bibnamefont {Shi}},\ and\ \bibinfo {author} {\bibfnamefont {J.}~\bibnamefont {Zi}},\ }\bibfield  {title} {\bibinfo {title} {Generating optical vortex beams by momentum-space polarization vortices centred at bound states in the continuum},\ }\href@noop {} {\bibfield  {journal} {\bibinfo  {journal} {Nat. Photonics}\ }\textbf {\bibinfo {volume} {14}},\ \bibinfo {pages} {623} (\bibinfo {year} {2020})}\BibitemShut {NoStop}%
\bibitem [{\citenamefont {Bomzon}\ \emph {et~al.}(2002)\citenamefont {Bomzon}, \citenamefont {Biener}, \citenamefont {Kleiner},\ and\ \citenamefont {Hasman}}]{RN18}%
  \BibitemOpen
  \bibfield  {author} {\bibinfo {author} {\bibfnamefont {Z.}~\bibnamefont {Bomzon}}, \bibinfo {author} {\bibfnamefont {G.}~\bibnamefont {Biener}}, \bibinfo {author} {\bibfnamefont {V.}~\bibnamefont {Kleiner}},\ and\ \bibinfo {author} {\bibfnamefont {E.}~\bibnamefont {Hasman}},\ }\bibfield  {title} {\bibinfo {title} {Space-variant pancharatnam-berry phase optical elements with computer-generated subwavelength gratings},\ }\href@noop {} {\bibfield  {journal} {\bibinfo  {journal} {Opt. Lett.}\ }\textbf {\bibinfo {volume} {27}},\ \bibinfo {pages} {1141} (\bibinfo {year} {2002})}\BibitemShut {NoStop}%
\bibitem [{\citenamefont {Shao}\ \emph {et~al.}(2018)\citenamefont {Shao}, \citenamefont {Zhu}, \citenamefont {Chen}, \citenamefont {Zhang},\ and\ \citenamefont {Yu}}]{RN41}%
  \BibitemOpen
  \bibfield  {author} {\bibinfo {author} {\bibfnamefont {Z.}~\bibnamefont {Shao}}, \bibinfo {author} {\bibfnamefont {J.}~\bibnamefont {Zhu}}, \bibinfo {author} {\bibfnamefont {Y.}~\bibnamefont {Chen}}, \bibinfo {author} {\bibfnamefont {Y.}~\bibnamefont {Zhang}},\ and\ \bibinfo {author} {\bibfnamefont {S.}~\bibnamefont {Yu}},\ }\bibfield  {title} {\bibinfo {title} {Spin-orbit interaction of light induced by transverse spin angular momentum engineering},\ }\href@noop {} {\bibfield  {journal} {\bibinfo  {journal} {Nat. Commun.}\ }\textbf {\bibinfo {volume} {9}} (\bibinfo {year} {2018})}\BibitemShut {NoStop}%
\bibitem [{\citenamefont {Fu}\ \emph {et~al.}(2020)\citenamefont {Fu}, \citenamefont {Zhai}, \citenamefont {Zhang}, \citenamefont {Liu}, \citenamefont {Song}, \citenamefont {Zhou},\ and\ \citenamefont {Gao}}]{RN37}%
  \BibitemOpen
  \bibfield  {author} {\bibinfo {author} {\bibfnamefont {S.}~\bibnamefont {Fu}}, \bibinfo {author} {\bibfnamefont {Y.}~\bibnamefont {Zhai}}, \bibinfo {author} {\bibfnamefont {J.}~\bibnamefont {Zhang}}, \bibinfo {author} {\bibfnamefont {X.}~\bibnamefont {Liu}}, \bibinfo {author} {\bibfnamefont {R.}~\bibnamefont {Song}}, \bibinfo {author} {\bibfnamefont {H.}~\bibnamefont {Zhou}},\ and\ \bibinfo {author} {\bibfnamefont {C.}~\bibnamefont {Gao}},\ }\bibfield  {title} {\bibinfo {title} {Universal orbital angular momentum spectrum analyzer for beams},\ }\href@noop {} {\bibfield  {journal} {\bibinfo  {journal} {Photonix}\ }\textbf {\bibinfo {volume} {1}} (\bibinfo {year} {2020})}\BibitemShut {NoStop}%
\end{thebibliography}%

\end{document}